\documentclass[11pt]{article}
\usepackage{bm,amsmath,color,dsfont, amssymb}
\usepackage[titletoc,toc,title]{appendix}
\usepackage{graphicx}
\usepackage[section]{placeins}
\usepackage{esint}
\usepackage{ulem}
\usepackage{adjustbox}
\usepackage[hidelinks]{hyperref}
\usepackage{caption}
\usepackage{cleveref}
\usepackage[round,authoryear]{natbib}
\usepackage{subcaption}
\usepackage{tabularx}
\usepackage{tgtermes} 
\usepackage{relsize}
\usepackage{booktabs}
\usepackage{wrapfig}
\usepackage{siunitx}
\usepackage{bm}
\usepackage{xcolor}
\usepackage{longtable}
\usepackage{booktabs}
\usepackage{tikz}
\usetikzlibrary{calc}
\usetikzlibrary{arrows, decorations.markings}
\usetikzlibrary{positioning}
\usetikzlibrary{decorations.pathreplacing}
\usetikzlibrary{arrows.meta}
\usetikzlibrary{shapes}
\usetikzlibrary{decorations.text}
\usetikzlibrary{math}
\usetikzlibrary{tikzmark}
\usetikzlibrary{backgrounds}
\usepackage{xcolor}
\definecolor{airforceblue}{RGB}{39, 80, 176}
\definecolor{orange1}{RGB}{255, 129, 3}
\definecolor{bordeau}{RGB}{167, 1, 69}

\usepackage{booktabs}
\usepackage{multirow}
\usepackage{float}

\newcommand{\Fref}[1]{Fig.~\ref{#1}}
\usepackage{authblk}
\title{Fragility-Resilience Trade-off and\\ Excess Volatility
in Supply Chain Networks}
\author[a]{David Martin}
\author[b,c,d]{Jos\'e Moran}
\author[e]{Debabrata Panja}
\author[f,g,h]{Jean-Philippe Bouchaud}

\affil[a]{LPTMC, Sorbonne Université, Paris}
\affil[b]{Macrocosm, Inc., Brooklyn, NY}
\affil[c]{Centre Borelli, ENS Paris-Saclay, Univ. Paris-Saclay, Paris, France}
\affil[d]{Institute for New Economic Thinking at the Oxford Martin School, University of Oxford, Oxford, UK}
\affil[e]{Department of Information and Computing Sciences, Utrecht University, The Netherlands}
\affil[f]{Capital Fund Management, Paris, France}
\affil[g]{X-CFM Chair of EconophysiX, Ecole polytechnique, Palaiseau, France}
\affil[h]{Académie des Sciences, Paris, France}

\begin{document}

\maketitle

\begin{abstract}
We study a production network in which firms use non-substitutable (Leontief) inputs, hold precautionary inventories and face idiosyncratic productivity shocks, with adjustment occurring through quantities rather than prices. We show analytically and numerically that a critical boundary exists in the space of shock volatility and inventory holdings: above this threshold, the economy absorbs shocks and fluctuates mildly while below it, cascading shortages make system-wide crises inevitable. 
Close to the threshold, aggregate output volatility diverges through network-mediated amplification of purely idiosyncratic shocks, providing a concrete mechanism for the ``small shocks, large business cycles'' puzzle. Because inventories are costly, competitive pressure drives firms toward the fragility boundary: a resilience-efficiency trade-off emerges, putting the gains from lean supply chains at risk. Finally, we show that supplier diversification shifts the threshold and, depending on its abundance, can eliminate the fragile regime entirely.
\end{abstract}

\tableofcontents

\section{Introduction and literature review}

When asked whether the US Federal Reserve had tools to deal with supply chain disruption and its consequences on inflation, Fed Chair Jerome Powell candidly admitted, ``{\it we don't have the kind of tools to deal with supply chain problems}''.\footnote{See \citet{CNN2025Powell}.} This highlights a blind spot in macroeconomic modelling: the lack of tractable frameworks for understanding and mitigating supply chain driven shocks. Over the past decade, a sequence of {exogenous shocks}, sometimes massive, sometimes minor --- the Covid-19 lockdowns, geopolitical tensions, and unpredictable events like the Ever Given container ship getting stuck in the Suez Canal --- have clearly exposed how fragile production networks can be \citep{nirei_carvahlo_2021, colon_scgovernance_2022}. These episodes have generated large and persistent macroeconomic fluctuations even when the underlying shocks were local, short-lived, or sector-specific.

Over several decades, global production has evolved towards leaner \textit{just-in-time} supply chains \citep{fujiwara_bankruptcy_2008, barrot_specificity_2016}. Firms have strong incentives to keep inventories as low as possible to minimize working-capital and storage costs, and to respond flexibly to changes in consumer demand. This tendency to run tight supply chains has driven operational costs down, but at the price of increased fragility, as discussed in \citet{hynes2022systemic}, \citet{Moran2025-xm}, and \citet{bouchaud2024self}. Empirical work has shown that for the case of France, firms that kept low inventories or had few alternative suppliers suffered substantially larger output losses during the Covid-19 supply chain disruption than firms that entered the crisis with higher inventories and more diversified suppliers, as shown by \citet{Lafrogne-Joussier2022-wv}.

From the perspective of macroeconomic modelling, this raises a natural question: how should we represent an economy in which production is organised through complex supply chains \citep{brintrup2015supply}, firms follow realistic inventory management policies, and small disruptions can snowball into large aggregate fluctuations? In a benchmark composed of independent agents or sectors, idiosyncratic shocks diversify away and aggregate volatility decays as $1/\sqrt{N}$ with the number of sectors $N$, in stark contrast with observed macro volatility.

To explain why large fluctuations persist at the aggregate level, several hypothetical mechanisms have been proposed. 
The first one relies on the existence of global shocks, which affect all firms and sectors simultaneously. 
Yet, when such shocks can be identified, they are often too small to account for the observed volatility of industrial production: \citet{Bernanke1994} coined it the ``small shocks, large cycles puzzle''. Another mechanism, dubbed the ``granularity hypothesis'', has been proposed by \citet{gabaix_Granular_origins}: the presence of ``mega-firms'', with outsized contributions to GDP, prevents idiosyncratic shocks from averaging out. While there is empirical support for this claim, other work suggests that propagation effects provide yet another mechanism which is at least as important \citep{mandel2025granular}.
This mechanism details how idiosyncratic shocks can actually cascade along the input--output network to become effectively macroeconomic, see e.g.\ \citet{fujiwara_bankruptcy_2008, barrot_specificity_2016, nirei_carvahlo_2021} and \citet{taschereau2025cascades}.

\paragraph{Equilibrium production network models.}
This mechanism of network-induced propagation has motivated a large literature on production networks with explicit inter-sectoral or inter-firm input--output linkages; see \citet{carvalho2019production} for a review. Building on the multisector framework of \citet{long_plosser_1983_rbc}, the influential contribution of \citet{acemoglu2012network} shows that when firms use Cobb--Douglas technologies in a network structure, the centrality of the network shapes the aggregate impact of idiosyncratic shocks, providing a network-based rationale for fat-tailed firm size distributions and their macroeconomic consequences \citep{carvalho2013great}. The Baqaee--Farhi framework \citep{baqaee_2018_cascading} generalizes this approach to Constant Elasticity of Substitution (CES) technologies, hence allowing for partial input substitution, showing that nonlinearities in the production structure can amplify shocks beyond what the Cobb--Douglas case predicts. 
These models have provided deep insights into how network topology mediates shock transmissions. 
These models all have in common the assumption of \textit{market clearing}: prices are taken to adjust so that supply equals demand in every market, whether the model is static or fully dynamic and stochastic. This rules out, by construction, the short-run phenomena that drive supply-chain fragility: stockouts, quantity rationing, delivery lags, finite and perishable inventories and the cascades of shortages they trigger. Many of these models are, in addition, analysed by comparing pre- and post-shock equilibria, which further abstracts from the transient dynamics through which these disruptions actually propagate. 
This assumption also implicitly requires price flexibility of an implausible magnitude. Using low short-run elasticities of substitution that are relevant for energy, clearing the market through prices alone would imply enormous prices increases: \citet{geerolf:hal-04015954} demonstrates that the Baqaee--Farhi approach applied by \cite{Bachmann2024} to the 2022 European gas crisis requires a 35-fold price increases to clear markets, whereas in practice adjustment occurred largely through rationing and demand destruction. As a result,  models that impose market-clearing are of limited use for understanding the type of disruptions observed during Covid-19,  natural disasters, or shipping blockages, which are patently disruptions that cause market-clearing to be violated. This also explains why they are incapable of capturing the disequilibrium effects on inflation caused by rationing during these shocks.\footnote{Another thing to note is that the identification of an elasticity of substitution with the parameter $\sigma$ in a CES function requires the assumption of competitive market clearing to hold.}

\paragraph{Network fragility and disruption cascades.}
A growing strand of work studies the emergence of fragility in production networks and how small shocks trigger large cascades. \citet{baqaee_2018_cascading} shows that with low input substitutability, firm-level shocks can cascade through the network and produce aggregate nonlinearities. \citet{colon2020fragmentation} examine the tension between  resilience and efficiency in  in the context of fragmentation and outsourcing, in which production processes are distributed across multiple entities. They find that, with more fragmentation, firms individually benefit from lower inventories, but this inadvertently leads to more extensive disruption cascades. \citet{taschereau2025cascades} models cascading firm shutdowns in endogenous production networks and characterizes how cascade sizes depend on network structure. \citet{Elliott2022} demonstrate that firms endogenously underinvest in the quality of their supply relationships, placing the equilibrium network on the verge of a ``precipice'' where aggregate reliability drops discontinuously -- a fragility threshold (or phase transition, in the language of statistical physics) driven by strategic underinvestment. \citet{Kopytov2024} show that supply chain uncertainty causes firms to choose safer but less productive suppliers -- a ``flight to safety'' that endogenously reshapes the production network. \citet{Elliott2025} provide a theoretical framework for the very short run in which physical availability of inputs is the binding constraint, with proportional rationing and price invariance determining real outcomes. \citet{Schaal2025} introduce heterogeneous production delays into the \citet{long_plosser_1983_rbc} framework and show that cycles in the production network generate oscillations -- ``echoes'' that travel along supply chains -- see also \cite{Dessertaine2022}. 
A common thread in this literature is the recognition that production networks can harbour sharp fragility thresholds.
However, most of these models neglect the role of inventories in absorbing or amplifying shocks -- excepting \citep{colon2017economic}.

\paragraph{Simulation-based supply chain models.}
In parallel, a growing empirical and simulation-based literature has focused on supply-chain disruptions and inventory dynamics. The ARIO models developed by \citet{Hallegatte2008-lq,Hallegatte2014-km} explicitly incorporate input inventories and have been used to simulate the economic consequences of natural disasters. \citet{Henriet2012} generalizes the ARIO framework to the level of individual firms interacting through a production network. 
They find that network concentration and clustering are key drivers of economic robustness -- a finding that points to a fundamental efficiency--resilience trade-off. 
Subsequent extensions, \citet{Inoue2019-ka} and \citet{colon2021criticality}, invariably show that the temporal profile and size of inventories strongly shape the depth and propagation of shocks, in line with the empirical evidence of \citet{Lafrogne-Joussier2022-wv} and established findings in the operations literature on demand amplification along supply chains (\textit{i.e.} the ``bullwhip effect'') \citep{sterman2006operational, brintrup2015supply}. 
More recent work by \citet{Pichler2022} confirms that calibrating inventory levels is crucial for reproducing the macroeconomic impact of Covid-19, and that inventory management policies were central to mitigating output losses during the 2022 Russian gas crisis; see also \citet{mandel2023disequilibrium}. These models are empirically rich and policy-relevant, but their realism comes at a cost: because they operate at very large scale with detailed data, they are difficult to analyze theoretically, and it is hard to derive clean conditions under which large cascades occur \citep{taschereau2025cascades}. A complementary strand of work, exemplified by \citet{Bonart2014} and \citet{Dessertaine2022}, has taken the opposite route: stripping production networks to a behavioral core to study their dynamics analytically. \citet{Dessertaine2022} find that the equilibrium solution of \citet{acemoglu2012network} can be dynamically unstable outside a certain parameter region, with large endogenous fluctuations emerging even without aggregate shocks.

\paragraph{Self-organised criticality and critical points in economics.}
The idea that the economy might be intrinsically unstable has a long pedigree. \citet{hawkins1948some} already raised this possibility, which was picked up again by \citet{bak1993aggregate} in the context of ``self-organised criticality'' in complex systems: a state in which small microscopic perturbations give rise to macroscopic fluctuations, like avalanches in sandpiles \citep{bak2013nature,bouchaud2024self}. \citet{Hallegatte2008} show, in a related vein, that introducing disequilibrium dynamics into a macroeconomic growth model can produce endogenous business cycles via a Hopf bifurcation --- a precursor to the nonlinear instabilities we study here. After remaining dormant for decades, the self-organised criticality scenario has attracted renewed attention \citep{bouchaud2024self}. \citet{moran2019may} show that firm network models {\it \`a la} Acemoglu et al.\ may generically evolve towards a critical point (or critical state) where a Hawkins-like transition occurs. \citet{Nirei2024} document that inflation dynamics exhibits repricing avalanches, with a broad distribution of shock sizes suggestive of proximity to such a critical point where inflation volatility would diverge (see \citet{bouchaud2024self} for a broader discussion). A more realistic incarnation of the \citet{bak1993aggregate} model has recently been proposed by \citet{Moran2024-bw,Moran2025-xm}, in the context of delay propagation in socio-technical systems such as train and flight networks \citep{Dekker2021}. In that model, there exists a critical point as the size of mitigating ``time buffers'' is reduced: if sufficient slack is built into schedules, delays can be absorbed; if not, delays accumulate and cause system-wide disruptions. This is clearly analogous to the inventory problem in supply chains: inventories are the natural counterpart of temporal buffers, since buffers ``buy time'' that allows production to proceed unimpeded during periods of upstream scarcity. Competitive pressure drives firms to reduce inventories and rely on just-in-time deliveries, pushing the system towards this critical point.

\paragraph{Disequilibrium economics and our contribution.}
Our contribution sits at the intersection of these four literatures. We propose a tractable dynamical model of production networks that makes inventories, quantity constraints, and network structure central, while establishing the existence of a resilience-to-fragility boundary -- analogous to a phase transition in statistical physics -- akin to what is observed in delay propagation models on complex networks \citep{Dekker2021,Moran2024-bw,Moran2025-xm}. 
In our setting, the analogue of time buffers is the amount of input goods that firms store to absorb upstream disturbances in the supply chain. 
For a given magnitude of productivity shocks, when precautionary inventories are too low the network becomes prone to system-wide disruptions, even in the absence of aggregate shocks. 
This provides theoretical foundations for the fragility patterns observed in ARIO-type simulations \citep{Hallegatte2008-lq,Henriet2012}, a disequilibrium complement to the equilibrium fragility threshold of \citet{Elliott2022}, and a concrete mechanism for the ``small shocks, large cycles'' puzzle.

Our model is deliberately stylised. Firms produce a single good using Leontief technologies with perishable, non-substitutable intermediate inputs; they face idiosyncratic productivity shocks and follow simple, myopic rules for adjusting production targets and orders. The key parameter is the depth of precautionary inventories $\kappa$: $\kappa = 0$ corresponds to just-in-time operation, while $\kappa = n$ means a firm holds enough inputs to sustain production for $n+1$ periods even if upstream supply is entirely cut off. 
We show that, despite its simplicity, this framework generates a sharp transition -- analogous to a phase transition in statistical physics -- between a resilient and a fragile, crisis-prone regime. 

The disequilibrium tradition that emerged in the 1960s and 1970s --- brought forth by \cite{clower1965keynesian,leijonhufvud1968keynesian,barro1971general,benassy1975neo,benassy2014macroeconomics} and many others --- made precisely this conceptual move: when prices fail to clear markets in the short run, agents face quantity constraints. The economy then switches between regimes depending on which constraints bind, as best exemplified by the contrast between \textit{Keynesian unemployment} --- where both goods and labour markets are in excess supply, so that firms cannot sell as much as they would like and therefore demand less labour than households are willing to provide --- and \textit{classical unemployment} --- where the goods market is in excess demand but the labour market remains in excess supply, so that firms are constrained on the supply side (for example if real wages are too high for full employment) rather than by insufficient demand. A third regime, \textit{repressed inflation}, arises when both markets are in excess demand. Each regime calls for different policy responses: Keynesian unemployment is alleviated by demand stimulus, while classical unemployment is not \citep{malinvaud1977theory}.

This disequilibrium-based program was abandoned not because it was empirically wrong~\citep{Backhouse_Boianovsky_2012}, but because (i) the dynamics across these regimes was analytically intractable in any model with more than two markets, and (ii) the emphasis in macroeconomics moved, after the critique by \cite{Lucas1976}, towards optimisation-based microfoundations, which stamped fixed price assumptions as ad-hoc [see \cite{howitt1979evaluating}] . 
Our model takes inspiration from this framework with two novel advantages: a \textit{network} of markets (exemplified by firms) rather than just two markets, and new \textit{computational and analytical} tools that resolve the tractability problem the original literature could not.

The disequilibrium tradition has seen a recent revival. Recent literature embeds quantity rationing in general-equilibrium models with monopolistic competition, recovering the regime-switching logic of the 1970s while gaining analytic tractability~\citep{Michaillat2015,barro1971general}. This is done by adding uncertainty and imperfect substitutability to smooth the discontinuities that made the original research programme difficult, allowing the authors to obtain a well-defined distribution of markets in excess supply or demand. Even more recently, \citet{Werning2026} have revisited the stability of the price-adjustment process itself, showing that forward-looking, rational price-setting can restore the convergence to equilibrium that the classical \textit{t\^atonnement} results had denied. We stress, however, that such an account remains \textit{equilibrium-like} in a specific and important sense. Its dynamics are written as differential equations in which the spillovers between rationed agents are resolved \textit{instantaneously}: the effective demand is a Leontief inverse $(I-A)^{-1}$, that is, a within-instant fixed point of the spillover map, so that between $t$ and $t+\mathrm{d}t$ the entire network is taken to have already settled into a mutually consistent configuration. Stability further requires that agents coordinate, from the outset, on a unique forward-looking path. Both features presuppose a degree of instantaneous coordination, with no representation of who observes which quantities, who acts first, or how long a disruption takes to propagate from one agent to the next.

Our approach is different, and deliberately so. Rather than assuming spillovers are resolved instantaneously, we specify an explicit, decentralised protocol --- namely a specification of which firm observes which quantities, when orders are placed, with what lag deliveries arrive, and how inventories are drawn down in the meantime---so that the propagation of shortfalls through the network \textit{is} the dynamics, rather than something assumed to have already converged. This is similar to the dynamics posited in \citet{Dessertaine2022}: the Leontief inverse is replaced by a temporal cascade that can be arrested, can accelerate, and can leave different firms in different rationing regimes at the same instant.

Our analysis is therefore firmly grounded in a quantity-rationing, disequilibrium setting as described above: in each period, markets need not clear and adjustment occurs through production, orders, and inventories rather than through prices. We interpret this as describing the ``very short run'' of \citet{Elliott2025}, in which physical availability of inputs is the binding constraint, and which can be highly relevant in practice, as we discuss in the next paragraph. Elliott and Jackson prove that in this regime, short-run impact is fundamentally different from equilibrium impact, and that proportional rationing fully determines real outcomes independently of prices---what they term \textit{price invariance}. Our model operates in this regime.

Several considerations support this modelling choice. First, the empirical elasticities of substitution between intermediate inputs are extremely low at short horizons. \citet{geerolf:hal-04015954} illustrates the consequences starkly for the 2022 European gas crisis: inverting the CES demand function with an elasticity of substitution $l\approx 0.1$ and a 30\% supply shortfall implies that market-clearing prices would need to rise roughly 35-fold---or about 14-fold for the broader energy sector with $l \approx 0.04$ and a 10\% cut. Prices of this magnitude are not observed in practice because demand destruction and physical rationing, not substitution, are the operative short-run mechanisms. Sticky long-term contracts, anti-gouging regulation, and reputational concerns further prevent prices from performing the allocative role that frictionless models assign them. Indeed, after supply disruptions, firms typically ration on the basis of existing customer relationships rather than willingness to pay, and information about the severity of imbalances propagates through direct observation of quantities---demand received, backlogs, inventory levels---rather than through price signals \citep{Hallegatte2008-lq,Hallegatte2014-km}. One may also view inventories as providing dynamic substitutability: short-run measured elasticities are near zero precisely because buffer stocks absorb fluctuations, converging to the true technological elasticity only over longer horizons. The Leontief production structure is therefore most descriptive in the short-run regime where inventories are the binding margin of adjustment.

Despite these considerations, we acknowledge that introducing price flexibility would not leave our results unchanged. 
In particular, if prices can rise in response to scarcity, they reduce demand before inventories are fully depleted, so the economy would tolerate a higher level of shock volatility $\sigma$ before becoming fragile---in other words, the critical boundary $\sigma_c(\kappa)$ would shift upward.
Therefore, our framework characterizes a worst-case benchmark in which the network must absorb shocks through quantities alone. 
More broadly, the trade-off between resiliency and efficiency is a recurring theme in supply chain literature: \citet{Kopytov2024} show that firms facing supply chain uncertainty may endogenously choose safer but less productive suppliers---a ``flight to safety'' that lowers expected GDP while reducing aggregate volatility. Our inventory mechanism captures an analogous trade-off at a different margin: larger buffers raise the critical shock threshold $\sigma_c$ but tie up resources. 
Even in the worst-case benchmark, the interaction between network structure, inventories, and shocks generates a sharp resilient-to-fragile transition along with excess volatility---mechanisms that would persist, though quantitatively modified, in a richer model with price adjustment and financial constraints. We discuss such extensions, together with policy implications, in the conclusion. 

Our specific contributions can be summarized as follows. We find:
\begin{itemize}
    \item[(i)] A well-defined transition line $\sigma_c(\kappa)$ in the space of shock volatility and inventory buffers that separates a stable stochastic steady state from a regime where system-wide crises become certain.
    \item[(ii)] A sharp increase and apparent divergence of aggregate volatility as this line is approached from above, even though shocks are purely idiosyncratic. The transition is continuous: aggregate volatility grows smoothly and diverges as inventories approach the critical level. Importantly, even past the critical point, the steady state remains {\it locally} stable -- it is large, cascading perturbations that destabilise the economy, producing negative skewness in the distribution of aggregate output.
    \item[(iii)] Strong effects of supplier diversification and rewiring capacity on the location and even existence of this critical boundary, provided firms can locate and switch to alternative suppliers fast enough relative to the pace at which shortages cascade.
    \item[(iv)] An endogenous mechanism through which competitive pressure on inventory costs and the mirage of stability can drive the economy toward the critical boundary, a scenario reminiscent of the famous Minsky cycles [\cite{minsky2008stabilizing}].
\end{itemize}

We characterize the transition both using numerical simulations and providing an analytical description in a high-connectivity, high-perishability limit (see section \ref{sec:HP-MF}). {A table of symbols used in the paper, together with their  definition, is provided at the end of the paper for convenience.}

\section{Set-up of the model}
 
We consider firms that specialise in producing a single product, i.e., firm $i$ produces good $i$. Every firm requires a certain number of intermediate inputs to produce its own good. The set of supplier firms needed for firm $i$ is denoted as $\{j\}\in{\cal S}(i)$. For definiteness, we will consider in this paper a random regular network of customers and suppliers, where each firm has exactly $K$ customers and $K$ suppliers randomly chosen within the universe of $N$ firms, with $K \ll N$ (other network topologies could also be envisioned). 

Firm $i$ keeps a stock of the goods $j$ needed to produce its product, denoted by $S_{ij}$, as well as an inventory of its own finished good, denoted by $g_i$. 

For simplicity, we assume that firm $i$ uses all goods $\{j\}\in{\cal S}(i)$ in equal capacity to produce a quantity $y_i$ of good $i$. We thus write the following {\it Leontief  production function} for good $i$ 
\begin{equation}
y_i = \min\left[y^\top_i,z_i\, \min_{j\in{\cal S}(i)} S_{ij},z_i L_0\right],
\label{e1}
\end{equation}
where $y^\top_i$ is the target production (see below), and $L_0$ is the amount of labour. The equation conveys that the production of good $i$ is lower limited by the lowest amount of supplied good $j$, and, similarly, by the availability of labour $L_0$. In \eqref{e1}, the productivity factor $z_i$ represents the efficiency of firm $i$: a higher $z_i$ leads to an increased production at fixed labour and supply. All quantities within \eqref{e1} are assumed to be positive or zero. Exogenous shocks will be modelled by a stochastic white noise affecting all $z_i$'s independently (a common noise factor is also possible, but our aim is to understand how idiosyncratic shocks can lead to system-wide  volatility). More precisely, we will write
\begin{equation}
    z_i(t) = z \,  e^{\xi_i(t) - \frac{\sigma^2}{2}},
\end{equation}
where $\xi_i(t)$ are i.i.d. Gaussian variables of zero mean and variance $\sigma^2$, and $z$ is the average productivity, assumed to be equal for all firms.

In all that follows, we will always consider $L_0$ to be large enough so that labour is never a binding constraint. This allows us to drop this term in the rest of our analysis.

\begin{figure}
\centering
\begin{tikzpicture}[
      >=Stealth, 
      node distance=1.2cm and 1.5cm,
      main/.style={circle, draw, minimum size=1cm, thick, font=\large},
      small/.style={circle, draw, minimum size=0.6cm, thick, font=\small, inner sep=1pt},
      rect/.style={rectangle, draw, minimum width=1.1cm, minimum height=0.6cm, thick, font=\small},
      mid arrow/.style={
          postaction={decorate},
          decoration={
              markings,
              mark=at position 0.85 with {\arrow{Stealth}}
            }
        },
      tiny/.style={circle, draw, minimum size=0.3cm, thick},
      consumption/.style={circle, draw, minimum size=0.2cm, densely dotted, thick, fill=gray!20},
      mid arrow cons/.style={
          densely dotted,
          thick,
          postaction={decorate},
          decoration={
              markings,
              mark=at position 0.85 with {\arrow{Stealth}}
            }
        },
      supplier/.style={teal, draw, thick,fill,fill opacity=0.2,text opacity=1},
      firmi/.style={airforceblue, draw, thick,fill,fill opacity=0.2,text opacity=1},
      firm/.style={dashed, thick, fill opacity=0.12,text opacity=1},
      background/.style={densely dotted, line width=1.3,fill opacity=0.06, draw opacity=0.06},
      customer/.style={bordeau, draw, thick,fill,fill opacity=0.2,text opacity=1},
      orderstyle/.style={densely dashed,line width=0.7,postaction={decorate},
          decoration={
              markings,
              mark=at position 0.55 with {\arrow{Latex[scale=1.4]}}
            }}
    ]

    \node[main,supplier,firm] (n1) {1};
    \node[main,supplier,firm] (n2) [right=4cm of n1] {2};

    \node (t1b) [right=0.4cm of n1] {};
    \node (t2a) [left=0.4cm of n2] {};

    \node[small,supplier] (g1) [below=0.3cm of n1] {$g_1$};
    \node[small,supplier] (g2) [below=0.3cm of n2] {$g_2$};
    \node[consumption] (c1) [right=0.7cm of g1] {c};
    \node[consumption] (c2) [left=0.7cm of g2] {c};

    \draw[mid arrow,supplier] (n1) -- (g1);
    \draw[mid arrow,supplier] (n2) -- (g2);
    \draw[mid arrow cons] (g1) -- (c1);
    \draw[mid arrow cons] (g2) -- (c2);



    \node[main,firmi,firm] (ni) at ($(n1)!0.5!(n2) + (0,-3.5)$) {firm $i$};
    \draw[orderstyle] (ni.east) to [out=15,in=-45] node[pos=0.4, sloped, below] {$O_{i2}$} (n2.east);
    \draw[orderstyle] (ni.west) to [out=165,in=-135] node[pos=0.4, sloped, below] {$O_{i1}$} (n1.west);

    \node[small,firmi] (ui1) [above left=0.3cm and 0.3cm of ni] {$S_{i1}$};
    \node[small,firmi] (ui2) [above right=0.3cm and 0.3cm of ni] {$S_{i2}$};

    \draw[mid arrow cons] (g1) -- (ui1) node[midway, sloped, below] {$X_{i1}$};
    \draw[mid arrow cons] (g2) -- (ui2) node[midway, sloped, below] {$X_{i2}$};

    \draw[mid arrow,firmi] (ui1) -- (ni);
    \draw[mid arrow,firmi] (ui2) -- (ni);

    \node[small,firmi] (gi) [below=0.3cm of ni] {$g_i$};
    \node[consumption] (ci) [right=0.7cm of gi] {c};

    \draw[mid arrow,firmi] (ni) -- (gi) node[midway, right] {$y_i$};
    \draw[mid arrow cons] (gi) -- (ci);


    \node[small,customer] (u3) [below left=5.5cm and 0.1cm of t1b] {$S_{3i}$};
    \node[small,customer] (u4) [below right=5.5cm and 0.1cm of t2a] {$S_{4i}$};

    \draw[mid arrow cons] (gi) -- (u3) node[midway, sloped, below] {$X_{3i}$};
    \draw[mid arrow cons] (gi) -- (u4) node[midway, sloped, below] {$X_{4i}$};

    \node[main,customer,firm] (b1) [below left=0.4cm and 0.4cm of u3] {3};
    \node[main,customer,firm] (b2) [below right=0.4cm and 0.4cm of u4] {4};

    \draw[mid arrow,customer] (u3) -- (b1);
    \draw[mid arrow,customer] (u4) -- (b2);

    \node[tiny, customer] (bt1) [above left=0.5cm and 0.2cm of b1] {};
    \node[tiny, customer] (bt2) [above right=0.5cm and 0.2cm of b2] {};
    \draw[mid arrow,customer] (bt1) -- (b1);
    \draw[mid arrow,customer] (bt2) -- (b2);

    \draw[orderstyle] (b2.north) to [out=90,in=-45] node[pos=0.4, sloped, above] {$O_{4i}$} (ni.east);
    \draw[orderstyle] (b1.north) to [out=90,in=-135] node[pos=0.4, sloped, above] {$O_{3i}$} (ni.west);

    \begin{scope}[on background layer]
      \node[rectangle,draw, minimum width=8cm, minimum height=2.2cm,rounded corners, dotted, line width=1.1,bordeau,fill,background] (colorcust) at ($(b1)!0.5!(b2) + (0,0.44)$) {};
      \node[bordeau,text width=1cm,align=center] at (colorcust.center) {customers $\mathcal{C}(i)$};
      \node[rectangle,draw, minimum width=8cm, minimum height=2.1cm,rounded corners, dotted, line width=1.1,teal,fill,background] (colorsup) at ($(n1)!0.5!(n2) + (0,-0.45)$) {};
      \node[teal,text width=1cm,align=center] at (colorsup.center) {suppliers $\mathcal{S}(i)$};
    \end{scope}

    \begin{scope}[shift={(8,0)}]
      \tikzmath{\msize = 0.2; \msizesup = 0.25; \ysep=0.2;}
      \node[main,customer,firm,minimum size=\msize cm] (leg1) at (0,0) {};
      \node[main,supplier,firm,minimum size=\msize cm] (leg2) [right=\ysep cm of leg1] {};
      \node[main,firmi,firm,minimum size=\msize cm] (leg3) [right=\ysep cm of leg2] {};
      \node (legfirm) [below=0.0cm of leg2] {Firms};
      \node[small,customer,minimum size=\msizesup cm] (legs1) at (2,0) {};
      \node[small,supplier,minimum size=\msizesup cm] (legs2) [right=\ysep cm of legs1] {};
      \node[small,firmi,minimum size=\msizesup cm] (legs3) [right=\ysep cm of legs2] {};
      \node [right=0.6cm of legfirm] {Storage};

      \node[consumption,minimum size=\msizesup cm] (legc) at (4.5,0) {};
      \node [below=0.0cm of legc] {consumption};

      \draw[mid arrow cons] (0,-1.2) -- +(1.2,0) node [midway, below,yshift=-1pt] {good flow};
      \draw[orderstyle] (2,-1.2) -- +(1.2,0) node [midway, below,yshift=-1pt] {order flow};
      \draw[mid arrow,firmi] (4,-1.2) -- +(1.2,0) node [midway, below,yshift=-1pt,black] {production};
    \end{scope}

    \begin{scope}[shift={(10.4,-2.2)}]
      \node (prod) at (0,0) {$y_i \leftarrow \min\left[y^\top_i,z_i\, \min_{j\in{\cal S}(i)} S_{ij},z_i L_0\right]$};

      \node (stock) at (0,-1) {$S_{ij} \leftarrow \left[S_{ij}-\frac{y_i}{z_i}+X_{ij}\right](1-\psi)$};

      \node (stock) at (0,-2) {$O_{ij} \leftarrow \left[(\kappa+1)\displaystyle{\frac{y^\top_i}{z_i}}-S_{ij}\right]^+ $};

      \node (stock) at (0,-3) {$M_{ij}\leftarrow \displaystyle{\frac{O_{ij}}{\sum_{i \in {\mathcal{C}}(j)}O_{ij}+c}} \left[y_j+g_j\right]$};

      \node (stock) at (0,-4) {$ X_{ij}\leftarrow \min\left[O_{ij},M_{ij}\right] $};

      \node (stock) at (0,-5) {$\displaystyle g_i \leftarrow \bigg[g_i+y_i-\sum_{j \in {\mathcal{C}}(i)}^K X_{ji}-C_i \bigg](1-\psi) $};

    \end{scope}

  \end{tikzpicture}
  \caption{Schematic representation of the supply-chain dynamics for firm $i$ in the case $K=2$ with the corresponding update rules and fluxes.}
  \label{fig:schematic_network}
\end{figure}

\subsection{The update rules}

The dynamical update rules below are mostly accounting balances, except for the target production $y^\top_i$, for which we use a reasonable constant gain learning rule based on current realised production. 

For example, the stock dynamics at firm $i$ then reads
\begin{equation}
S_{ij}(t+1)=\left[S_{ij}(t)-\frac{y_i(t)}{z_i(t)}+X_{ij}(t)\right](1-\psi_j).
\label{e2}
\end{equation}
Equation \eqref{e2} states that the stock of good $j$ at firm $i$ and time $t+1$ is the stock of good $j$ at firm $i$ at time $t$ minus what the firm used up to produce good $i$ plus the quantity $X_{ij}(t)$ it has bought from firm $j$, all multiplied by an overall decay factor $(1-\psi_j)$, where $\psi_j$ is the perishability of good $j$. For durable goods, $\psi_j \approx 0$ and for instantaneously perishable goods, $\psi_j=1$. In the following, we will assume that $\psi_j \equiv \psi$ for all $j$. 

We now detail the equation for $X_{ij}$, the quantity bought by firm $i$ to firm $j$. 
It results from balancing the orders of firm $i$ with respect to the total delivery of firm $j$, which has to cater for the entire demand.
There are two types of buyers of good $j$. The first are the downstream firms, which seek to acquire the inputs needed for their planned production plus a safety buffer.
Thus, the orders of downstream firm $i$ to firm $j$ are given by:\footnote{Throughout this paper we will use the notation $[x]^+ = \max(x,0)$.}
\begin{equation}
O_{ij}(t)=\left[(\kappa+1)\frac{y^\top_i(t)}{z_i}-S_{ij}(t)\right]^+,
\label{e3}
\end{equation}
where the coefficient $\kappa \geq 0$ quantifies the safety ``buffer'': if $\kappa = 1$ then firms $i$ order twice the needed quantity.
In a benchmark where firms face no friction and no uncertainty, firms would hold no precautionary buffer ($\kappa=0$) and keep just enough input goods to produce their current target $y_i=z_i S_{ij}\quad \forall (i,j)$. Positive buffers are a response to shortfall risk, following a rationale we derive in Section~\ref{sec:soc_endogenous}.

Eq.~\eqref{e3} is known in the literature as an \textit{order-up-to} or \textit{base-stock} inventory policy. At each period, firm $i$ tops up its stock of input $j$ back to a target level $\displaystyle{\frac{\kappa+1}{z_i}}y_i^\top$. This rule is optimal when holding stock and running short both carry a per-period cost but placing an order carries no fixed cost; the result was obtained for a single firm by~\citet{Arrow1951}, see also~\citet{Zipkin2000-yk} for a textbook treatment. It is derived for a firm facing an exogenous shortfall-risk distribution. In our framework that distribution is endogenous to the entire economy, but we adopt the order-up-to policy as a natural benchmark. Its only free parameter is the size of the target, set by $\kappa$, which we fix endogenously in Section~\ref{sec:soc_endogenous} by weighing the cost of holding stock against the cost of running short.

When ordering carries a fixed cost, the optimal policy is instead of the $(s,S)$ type -- stock is allowed to drop to a lower threshold $s$ before a single large order refills it to an upper level $S$, generating lumpy, infrequent orders. This is the rule typically emphasised in the macroeconomics of inventories \citep{kahn1987,blinder_maccini_1991,caballero_engel_1991,caplin_1985,fisher_hornstein_2000,khan_thomas_2007}. The two rules aim at the \textit{same} target stock and differ only in \textit{when} orders arrive: the average stock that the data report---the inventory-to-sales ratio---is the same under both \citep{bils_kahn_2000}, whereas the lumpiness of individual orders, which our rule does not reproduce, is the distinctive feature of $(s,S)$~\citep{alessandria_2010}.

The other buyers of good $j$ are households outside the production system, which order a fixed amount $c_j$ each period. Firm $j$'s available output is then shared between its downstream firms and these households in proportion to their orders.
Thus, firm $i$ and households are respectively entitled to receive an amount of good $j$ equal to
\begin{equation}
M_{ij}(t)=\displaystyle{\frac{O_{ij}(t)}{\sum_{k \in {\mathcal{C}}(j)}O_{kj}(t)+c_j}} \left[y_j(t)+g_j(t)\right]\quad\mbox{and}\quad m_j(t)=\displaystyle{\frac{c_j}{\sum_{k \in {\mathcal{C}}(j)}O_{kj}(t)+c_j}}\left[y_j(t)+g_j(t)\right],
\label{e4}
\end{equation}
where ${\mathcal{C}}(j)$ is the set of customers of $j$ and $g_j(t)$ its current stock of good $j$.
This corresponds to a \textit{relative} or \textit{proportional} rationing rule of \cite{benassy1982economics,benassy1975neo}, also treated axiomatically by \cite{drezeExistenceExchangeEquilibrium1975}. This choice matters: alternative rules such as queue rationing or priority-based allocations would generate different cascade dynamics, as recently explored by \citet{han2025dynamicshockrecoveryio}.

Finally, the actual amount of transacted goods is given by the smallest of the two quantities: either the raw order or the fair share in \eqref{e4}. 
This yields:
\begin{equation}
X_{ij}(t)=\min\left[O_{ij}(t),M_{ij}(t)\right]\quad\mbox{and}\quad C_j(t)=\min[c_j,m_j(t)].
\label{e5}
\end{equation}
Equations \eqref{e2} and \eqref{e5} then allow one to obtain $S_{ij}(t+1)$.

Similarly to \eqref{e2}, the update rule for the inventory $g_i(t)$ is obtained by subtracting the outbound flux of goods to its inbound counterpart, while taking into account the perishability. It thus reads
\begin{equation}
g_i(t+1)=\left[g_i(t)+y_i(t)-\sum_{j \in {\mathcal{C}}(i)}^K X_{ji}(t)-C_i(t)\right](1-\psi).
\label{e6}
\end{equation}
Note that, by construction, the right-hand side is always non-negative. 

The system of equations \eqref{e1}-\eqref{e2} and \eqref{e5}-\eqref{e6} yields a complete dynamical description of the variables \{$y_i$, $g_i$, $X_{ij}$, $S_{ij}$\}, provided we specify the target production $y^\top_i$. 
We will assume that $y^\top_i$ adapts to the current market situation at a constant learning rate. 
Firm $i$ observes the total demand for its good at the previous time step, corrected by its unsold stock.
It also measures the actual production it could have achieved with the input goods at its disposal. Using these two pieces of information, it updates its next production target as
\begin{equation}
\begin{split}
y^\top_i(t+1)= &(1 - \omega) y^\top_i(t) \\& + \omega \left[\min\left(\left[c_i + \sum_{j \in {\mathcal{C}}(i)}O_{ji}(t)-g_i(t)\right]^+,z_i(t)\min_{j\in \mathcal{S}(i)}S_{ij}(t),z_i(t)L_0\right)\right]\;.
\end{split}
\label{e7}
\end{equation}
In other words, firm $i$ attempts to anticipate the demand for its good and its ability to produce by adapting its production target at learning rate $\omega$.\footnote{This constant-gain rule is an exponential-smoothing demand forecast. Pairing order-up-to inventory policies with demand forecasting is the canonical source of the ``bullwhip effect'', or the amplification of order variance moving upstream~\citep{lee_1997_bullwhip}.}

The above dynamical updates are a mix of accounting rules (Eqs. \eqref{e2}, \eqref{e5}, \eqref{e6}) with a standard order-up-to inventory rule for anticipating supply shortfalls (Eq. \eqref{e3} with $\kappa > 0$) and a constant-gain adjustment of production targets [Eq. \eqref{e7}]. We work with the order-up-to-rule for two reasons. The first is that it is analytically tractable. The second one, which is more important, is that it isolates the cascade mechanism we want to study from the separate question of order \textit{lumpiness}. In our model, the transition is governed by \textit{expected} flows, as defined by the worst-supplier ratio $r_<$, the rate $\gamma$ and the rationing rule, rather than by the smoothness of any individual ordering. Lumpy $(s,S)$ ordering would reinforce the effect rather than remove it: \cite{caplin_1985} shows that these policies make the variance of orders \textit{exceed} the variance of sales, so the propagation of shortages along the supply chain survives and is probably strenghtened by lumpy ordering.

 Note that prices are not featured in the dynamics. As discussed in the Introduction, the Leontief production structure and the ``very short run'' time horizon we consider make this a natural modelling choice: under proportional rationing, real outcomes are independent of prices \citep{Elliott2025}, and the empirical elasticities of substitution at short horizons are too low for price adjustment to clear markets \citep{geerolf:hal-04015954}. We assume throughout that firms are homogeneous, sharing the same average productivity $z$, number of suppliers and clients $K$, and household target consumption $c$. 
We expect heterogeneity to sharpen rather than soften the mechanism highlighted here. A network with a fat-tailed degree or size distribution would concentrate flows on a few hubs, which then act as systemic bottlenecks, much as in the granularity and network-centrality literature~\citep{gabaix_Granular_origins,acemoglu2012network}. A more complete analysis (along the lines of \citealp{Dessertaine2022}) would be an interesting extension that we discuss in the conclusion.

\subsection{Stationary state}
\label{sec:stationary_state}

Let us now look at the stationary state of the model in the absence of shocks ($\sigma=0$), assuming that precautionary inventories are large enough to prevent productions from being limited by the availability of input goods. In this case, production is always sufficient to satisfy all needs, hence $X_{ij}=O_{ij}$, $y_i = y_i^\top$ and the homogeneous stationary state must satisfy the following equations:
\begin{equation}
\label{eq:steady_state_eq1}
   S^\star = \frac{1- \psi}{\psi} \left(X^\star - \frac{y^\star}{z} \right), \quad 
    X^\star = O^\star = (\kappa + 1) \frac{y^\star}{z} - S^\star,\quad 
    g^\star = \frac{1- \psi}{\psi}  \left( y^\star - K X^\star - c \right),
\end{equation}
with
\begin{equation}
    y^\star = c + K O^\star - g^\star = c + K X^\star - g^\star.
\end{equation}
The last two equations immediately lead to $g^\star = 0$ (no unsold goods), provided $\psi \neq 1$. We get
\begin{align}
\label{eq:eqproduction}
    X^\star =& \frac{(1 + \kappa \psi)\,c}{z - K (1 + \kappa \psi)}, & S^\star =&\frac{\kappa c\,(1-\psi)}{z-K(1+\kappa\psi)}, & y^\star = & \frac{zc}{z - K (1 + \kappa \psi)}\,.
\end{align}
Note that household consumption $c$ appears as a global scaling factor of production and inventories and will play a minor role in the following. For this market-clearing equilibrium steady state to exist, $\kappa$ must be bounded as $\kappa_{{\rm min}} < \kappa < \kappa^{\star}_c$, where $\kappa^{\star}_c$ and $\kappa_{{\rm min}}$ are given by 
\begin{align}
    \kappa^{\star}_c =& (z-K)(K\psi)^{-1}\;, & \kappa_{{\rm min}}=(1-\psi)^{-1}\;.
\end{align}

The first condition, namely $\kappa < \kappa^{\star}_c$, can be recast into
\begin{equation}
    \label{eq:ss_cond_1}
    z > K (1 + \kappa \psi)\;,
\end{equation}
which highlights that productivity must be sufficient to sustain the economy, including the inventory buffers. This is the dynamic equivalent of the Hawkins--Simon condition.

Likewise, the second condition, namely $  \kappa_{{\rm min}}< \kappa $, is expressed as
\begin{equation}
    \label{eq:ss_cond_2}
    \kappa (1-\psi)>1\;, 
\end{equation}
which highlights that more inventories must be stored if goods depreciate fast.
Finally, one can check that steady state \eqref{eq:eqproduction} always  satisfies Eq. \eqref{e5} since $M^\star = O^\star$.

\section{Stability against uniform perturbations}
\label{sec:stability}

Before turning to the role of idiosyncratic shocks, we need to verify that the steady state of Section~\ref{sec:stationary_state} is dynamically stable in the absence of shocks: our intention here is to distinguish deterministic instabilities from cascades driven by volatility. We outline the logic here and detail the corresponding algebra in Appendix~\ref{app:cone-wise}.

In all that follows, we write for any time-series $x_i(t) = x_i^\star + \Delta x_i(t)$, where now $\Delta x(t)$ indicates the deviation from the stationary-state value. Similarly, we drop all indices and consider purely homogeneous perturbations around the homogeneous steady state: since $x_i^\star=x^\star$ we study perturbations $\Delta x_i(t) = \Delta x(t)$.

\subsection{Two regimes: demand-limited and supply-limited}

Linearising the dynamics around the homogeneous steady state requires care because the system behaves differently depending on which constraint binds. When realised supply suffices to meet orders, $O(t) < M(t)$, the rationing in Eq.~\eqref{e5} is inactive and firms produce to meet target. When orders exceed available supply, $O(t) > M(t)$, the rationing rule binds, deliveries fall short of orders, and inventories at downstream firms are drawn down faster than they replenish. These two regimes correspond to two \textit{cones} in the space of perturbations $(\Delta a, \Delta y)$, where 

\begin{equation}
    \begin{split}
    \Delta a :=& \Delta y + \Delta g - K\Delta O=\left(1  - K\frac{\kappa+1}{z}\right) \Delta y+\Delta g+K\Delta S
    \end{split}
\end{equation}
captures the gap between realised production and desired (or notional) demand. 
The first two terms capture the production and the finished goods inventories, while the last term $K\Delta O$ captures orders.

The two regions $\Delta a>0$ and $\Delta a <0$ are the network analogues of \cite{malinvaud1977theory}'s regime classification. In the demand-limited cone $\Delta a>0$, each firm could in principle produce more if demand were higher: this is the analogue of \textit{Keynesian unemployment} in a single good model, but generalised to every node of the production network. In the supply-limited cone $\Delta a<0$ firms cannot produce as much as they would like because intermediate inputs are unavailable: this is the analogue of \textit{classical unemployment} (or, to be more precise, of \textit{repressed inflation} from the producer side, with the missing input as the binding scarcity). The novelty here is that an economy with $N$ firms can in principle be in $2^N$ possible regimes, with different firms occupying different cones ($\Delta a_i \gtrless 0$ for each $i$). The regime-switching dynamics that defeated analytical disequilibrium theory in the 1970s~\citep{howitt1979evaluating} are exactly what we resolve numerically --- and analytically in the high-perishability limit we present in following sections.

\subsection{Stability region}

\begin{figure}
    \centering
    \includegraphics[width=1.0\linewidth]{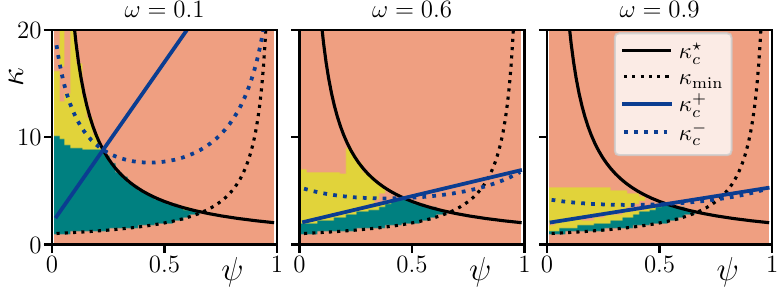}
    \caption{Numerical stability analysis of the model in the $(\psi, \kappa)$ plane for different values of $\omega$ and $K/z=1/3$. The different colors indicate unstable (coral), stable (teal) and endogenous oscillation (sand) regimes. We also show the four critical values of $\kappa$: $\kappa_{\min}, \kappa_c^\star, \kappa_c^-, \kappa_c^+$.}
    \label{fig:unstability_phase_diagram}
\end{figure}

In this section, we decompose the dynamics on the different cones and study the linear stability of the steady state within each cone. Starting from the exact homogeneous dynamics given by
\begin{equation}
\left[\begin{matrix}
    \Delta y(t+1) \\ \Delta a(t+1)
\end{matrix}\right] = f\left[\begin{matrix}
    \Delta y(t) \\ \Delta a(t)
\end{matrix}\right],
\end{equation}
where $f$ is a function retaining all the non-linearities in the model, we then linearise it within each cone as 
\begin{equation}
\label{eq:cone_wise_linear}
\left[\begin{matrix}
    \Delta y(t+1) \\ \Delta a(t+1)
\end{matrix}\right] = \mathcal{L}^\pm \left[\begin{matrix}
    \Delta y(t) \\ \Delta a(t)
\end{matrix}\right],
\end{equation}
where $\mathcal{L}^\pm$ are matrices governing the dynamics within the positive ($\Delta a > 0$) and negative ($\Delta a < 0$) cones. 
The expression of $\mathcal{L}^\pm$ can be respectively obtained by a Taylor expansion of $f$ around the steady state and eq \eqref{eq:cone_wise_linear} corresponds to the cone-wise linearisation~\citep{Dessertaine2022, dessertaine2022non}, which may be studied by finding the eigenvalues of $\mathcal{L}^\pm$. 
Eigenvalues with a real part larger than one are unstable, while those with a real part smaller than one are stable. Similarly, the presence of imaginary parts indicates the presence of endogenous oscillations. 
The two linearised operators $\mathcal{L}^+$ and $\mathcal{L}^-$ are given in Appendix~\ref{app:cone-wise}. Our analysis details the emergence of two characteristic quantities governing the stability of the steady-state solution at fixed $\omega$ and $\psi$.
We dubbed them $\kappa_{c}^{\pm}$ and their expression is given by 
\begin{align} \label{eq:kappas}
     \kappa_c^+ &:= \frac{z(\omega + \psi)}{K\omega} - 1, & \kappa_c^- &:= \frac{z (1+ \omega) - K ( 1 + \omega - \psi)}{K(\omega + \psi - \psi^2)},
\end{align}
As illustrated in Fig~\ref{fig:unstability_phase_diagram},
\begin{itemize}
    \item When $\kappa < \kappa_{\min}$ or $\kappa > \kappa^{\star}_c$, the steady state is not well defined, either because precautionary inventories are too small relative to the perishability rate or because productivity is too low, as shown in Eq.~\eqref{eq:ss_cond_2} and Eq.~\eqref{eq:ss_cond_1}.
    \item For $\kappa_{\min} < \kappa < \min(\kappa_c^\star, \kappa_c^+)$, both cones are linearly stable: small homogeneous perturbations decay back to the steady state.
    \item When $\max(\kappa_c^+,\kappa_{\min})<\kappa<\min(\kappa_c^-,\kappa_c^{\star})$, \textit{i.e.} in the intermediate range delimited by $\kappa_c^+$ and $\kappa_c^-$, one cone is locally unstable while the other is stable. Numerical simulations reveal that the system can undergo a Hopf bifurcation and give rise to persistent endogenous oscillations, which can be interpreted as trajectory switches from the unstable to the stable cone. 
    \item For $\kappa>\kappa_c^{-}$, both cones are unstable.
\end{itemize}
Note also that, as detailed in Appendix~\ref{app:cone-wise}, analytical results can be obtained in the limit $\omega \to 0$ and $\psi = O(\omega)$: the steady-state solution always remain stable in this regime.

\section{Excess Volatility and Crises: A Numerical Analysis}
\label{sec:numerical_analysis}

In the remainder of the paper, we focus on the linearly-stable region, defined by
\begin{equation} \label{eq:stable_region}
    \kappa_{\min} < \kappa < \min(\kappa_c^\star, \kappa_c^+),
\end{equation}
so that any observed instability must be entirely driven by idiosyncratic productivity shocks propagating through the supply chain: volatility cannot be attributed to the underlying deterministic dynamics. 

We simulate the dynamics defined in Eqs.~\eqref{e3}-\eqref{e7} on a directed, random regular network of $N$ firms, each with $K$ suppliers and $K$ customers. We define a \textit{crash} as a self-reinforcing downward spiral in which all firms reach a production level below $10^{-10}$ and never recover. Our baseline parameters are $\kappa_{\min} = 1.1$,  $\kappa_c^\star= 20$, $\kappa_c^+ = 5$, $\kappa_c^- = 12.1$ and $\kappa = 2.6$: we work well within the stable region.

Note that this is a modelling device for an absorbing collapsed state. Economically, it should be interpreted as a severe, self-sustaining contraction that the model's rules cannot escape. In reality, firm entry, policy intervention and input substitution would eventually stop such a spiral. We discuss such possibilities, in the form of rapid supplier rewiring, in Section~\ref{ref:hp-diversification}.

\subsection{The resilience-to-fragility transition}

Intuitively, one expects that there should be a transition line in the ($\sigma, \kappa$) plane separating a ``north-west'' (high inventories $\kappa$, low volatility of shocks $\sigma$) region where the economy is stable, but possibly volatile, from a ``south-east'' (low inventories, high volatility of shocks) region where production crashes because, in our non-substitutable Leontief economy, lack of input products results in a self-reinforcing downward spiral. 

\begin{figure}
    \centering
    \begin{tikzpicture}
        \node at (0,0) {\includegraphics{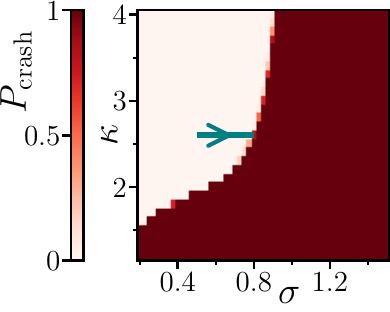}};
        \node[draw, densely dotted,line width=1.1, rounded corners, inner sep=10pt] (ev) at (8,0) {\includegraphics[width=5cm]{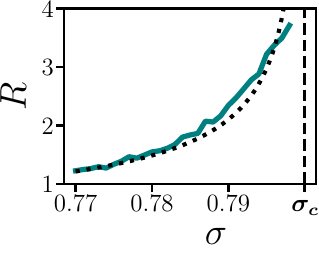}};
        \node[draw, minimum width=2cm, densely dotted, line width=1.5, minimum height=0.8cm,rounded corners] (spot) at (0.65,0.35) {};
        \draw[densely dotted,line width=1.1] (spot.north east) -- (ev.north west);
        \draw[densely dotted,line width=1.1] (spot.south east) -- (ev.south west);

        \node[scale=1.25] at (-0.6,2.1) {a.};
        \node[scale=1.25] at ($(ev.center)+(-1.1,1.5)$) {b.};

    \end{tikzpicture}
    \caption{{\bf a.} Heat map of the crash probability $P_{\rm crash}$ in the $(\kappa,\sigma)$ plane. Upon increasing $\sigma$, the economy transitions from resilient to fragile beyond a critical $\sigma_c(\kappa)$. For $(\sigma, \kappa)$ below the boundary line, the economy crashes almost surely. Note that there also appears to be a value $\sigma_{\max} \approx 1.$ beyond which the economy is always unstable, even when inventories are very high. {\bf b.} Excess Volatility $EV$ as a function of $\sigma$ for a certain value $\kappa$, such that $\sigma_c \approx 0.8$ (black dashed line). The dotted line is a fit as $R\propto \frac{1}{\sqrt{\sigma_c - \sigma}}$. {\bf Parameters:} $c=6$, $K=6$, $N=750$, $\omega=0.1$, $\psi=0.1$, $T=2000$, $z=18$, $L_0=1.$}
    \label{fig:phase_diagram}
\end{figure}

This is indeed confirmed by numerical simulations, as shown on Fig. \ref{fig:phase_diagram}a which displays a sharp boundary $\sigma_c(\kappa)$ above which crashes (in the context of the model) become almost certain. 
The boundary shifts upward with $\kappa$, implying that larger inventory buffers raise the shock amplitude the economy can tolerate.
However, the improvement in resilience ceases at a maximum value $\sigma_{\max} \approx 1.$ beyond which crashes occur regardless of the size of inventory buffers: we will recover this upper bound analytically in Section~\ref{sec:HP-MF}.

\subsection{Continuity of the transition and excess volatility}

Two features of this resilient-to-fragile transition matter for an economic interpretation.
First, it is a \textit{continuous} transition: the probability to observe a crash in a given time window rises smoothly from zero to one as $\sigma$ crosses the critical value $\sigma_c(\kappa)$, with no discrete jump. 
We verify this continuous nature by a finite-size scaling analysis, done in Appendix~\ref{app:finite-size}, where we follow the approach of \citet{fosset2020endogenous} for a model of liquidity crises in financial markets. 
The analysis shows that in the double limit $N,T \to \infty$ with $T\ll N^{3/2}$ the probability $\mathbb{P}_N[\tau_c \leq T]$ to observe a crash in a simulation of length $T$ converges to an interpolating function at $\sigma_c$, with a transition width shrinking as $1/\log^2(N)$. 
The exponents and scaling functions might depend on parameters such as the network's topology, but the qualitative nature of the transition --- continuous, and sharpening with system size --- is robust. 
This is a property known as \textit{universality} in statistical physics: the qualitative nature of the transition depends only on a small number of structural features and is insensitive to microscopic details. The existence and shape of this transition reflect structural properties of the underlying dynamics, not the particular behavioral rules we have chosen, and we therefore expect them to survive in richer environments.

The second economically interpretable feature is the divergence of aggregate production's volatility as the critical boundary is approached from below. We quantified this divergence with the excess volatility ratio $R$, defined as
\begin{equation}\label{eq:excess_vol_app}
    R^2:= \frac{\mathbb{V}[\sum_i y_i ] / \mathbb{E}[\sum_i y_i ]^2}{\mathbb{V}[\sum_i z_i ] / \mathbb{E}[\sum_i z_i ]^2} \;.
\end{equation}

The quantity $R$ has a natural economic interpretation: it measures how much aggregate output volatility the network produces \textit{per unit} of underlying idiosyncratic productivity volatility. In a represen\-tative-agent or fully-diversified economy, where interactions don't play a strong role, the two volatilities coincide and $R=1$. More generally, $R$ captures the extent to which network propagation amplifies ($R>1$) or dampens ($R<1$) the microeconomic shocks that drive it. This is the central finding of this section: idiosyncratic shocks of fixed amplitude $\sigma$ generate arbitrarily large aggregate volatility when $\sigma$ approaches the fragility boundary $\sigma_c$, providing a concrete mechanism for the ``small shocks, large cycles'' puzzle. Unlike~\cite{gabaix_Granular_origins}'s mechanism, which requires fat-tails in the firm-size distribution, the amplification here arises purely from network-mediated comovement between firms that are all of comparable sizes.

Equivalently, since aggregate productivity volatility scales as $\sqrt{1/N}$ by the law of large numbers (each firm's productivity shock is independent), so does aggregate output volatility whenever $R$ remains bounded --- the standard diversification argument. A divergence of $R$ therefore signals the breakdown of this diversification: aggregate output volatility no longer decreases with the size of the economy, and idiosyncratic shocks do not cancel out at the aggregate level.

Fig~\ref{fig:phase_diagram} shows that $R$ rises sharply as $\sigma\to \sigma_c(\kappa)$, and in fact diverges as $(\sigma_c(\kappa)-\sigma)^{-1/2}$. This is the central economic finding of this section.

\section{Analytical Results: The High-Perishability, Large-Economy Limit}
\label{sec:HP-MF}

While numerical results of section~\ref{sec:numerical_analysis} and Appendix~\ref{app:finite-size} establish that a resilient-to-fragile transition exists, they leave three important questions unanswered: What determines the location of the critical line $\sigma_c(\kappa)$? How does it depend on the structural parameters of the production network --- the number of suppliers, the depth of inventories, the productivity of firms? And lastly, what would shift the boundary in a way that policy or firm strategy could exploit? 

In this section, we answer these questions by studying a tractable limit of our model in which intermediate goods are highly perishable. 
The closed-form expression we obtain for $\sigma_c$ in this limit recovers the general numerical phase diagram and matches the upper bound $\sigma_{\max}$ identified in Section~\ref{sec:numerical_analysis}. 

\subsection{The scaling-consistent high-perishability limit}

To study our supply chain analytically, we consider the limit $\psi=1-\varepsilon$ with $\varepsilon \to 0$, in which intermediate goods perish almost immediately at each period. 
This limit describes a real economic regime for some goods: sectors providing electricity, fresh food, or time-sensitive logistics, where inventories cannot be stored across many periods. In this regime, the inventory variables $S_{ij}$ and $g_i$ become algebraically dependent to the flows (see Appendix~\ref{app:high-perishability}), so the only persistent dynamics in the model is the propagation through orders --- thereby isolating the cascade mechanism from inventory drift. 
This limit is analytically tractable while remaining qualitatively consistent with the general finite-perishability case, as our numerical comparison will confirm.

A non-trivial step is, however, required: taking the limit \textit{consistently} in a way that keeps the steady-state inside the linearly stable region defined in Section~\ref{sec:stability}. 
To remain in this region, the Hawkins-Simon condition \eqref{eq:ss_cond_1} requires $z>K(1+\kappa\psi)$, while the inventory feasibility condition \eqref{eq:ss_cond_2} requires $\kappa(1-\psi)>1$. As $\psi\to 1$, both bounds shift: inventories and productivity must scale up to remain adequate.

We therefore enforce the following scaling,
\begin{equation}
\label{eq:HP_scaling}
\psi = 1 - \varepsilon, \qquad \kappa=\frac{\varkappa}{\varepsilon}, \qquad 
z=\frac{K\zeta_0}{\varepsilon}
\end{equation}
with $\varkappa > 1$ and $\zeta_0 > \varkappa$. Under this rescaling, the steady-state production, orders and rescaled inventories remain of order one in the limit $\varepsilon \to 0$:

\begin{equation}\label{eq:HP_steady_state}
\begin{split}
    y^\star = \frac{c\zeta_0}{\zeta_0 - \varkappa}, \qquad 
    X^\star = \frac{c}{K}\frac{\varkappa}{\zeta_0 - \varkappa}, \qquad 
    z S^\star = \frac{c\varkappa\zeta_0}{\zeta_0 - \varkappa},
\end{split}
\end{equation}
Note however that there is a second stationary state where production has collapsed, i.e. $y=g=S=X=O=0$, towards which the system will evolve when productivity shocks are large enough.

Let us also compute the boundaries $\kappa_c^+$ and $\kappa_c^-$ in such a regime. From Eq. \eqref{eq:kappas} we find, for $\varepsilon \to 0$,
\begin{equation} \label{eq:kappas_2}
    \kappa_{\min} = \frac{1}{\varepsilon}, \qquad \kappa_c^+ \approx \frac{\zeta_0(1 + 1/\omega)}{\varepsilon}, \qquad \kappa_c^- \approx \kappa_c^+,
\end{equation}
and finally $\kappa_c^\star \approx \zeta_0/\varepsilon < \kappa_c^+$. Hence the linearly stable region defined by Eq. \eqref{eq:stable_region} reads here
\begin{equation} \label{eq:stable_region_2}
    1 < \varkappa < \varepsilon\min(\kappa_c^\star, \kappa_c^+) \equiv \zeta_0.
\end{equation}

In what follows, $\varkappa$ plays the role of a rescaled inventory buffer, while $\zeta_0$ plays the role of a rescaled productivity. The two parameters of interest for the resilient-to-fragile transition become $\varkappa$ and the shock amplitude $\sigma$: our task is then to study the critical value $\sigma_c(\varkappa)$ in the linearly stable regime $1 < \varkappa < \zeta_0$.

\subsection{Adding shocks: numerical phase diagram}
\label{sec:HP-numerical}

We now reintroduce idiosyncratic productivity shocks: each firm's productivity fluctuates around a common mean as:

\begin{equation}
    \label{eq:HP_productivity}
    z_i(t) := \frac{K\zeta_0}{\varepsilon} e^{\xi_i(t) - \sigma^2/2},
\end{equation}
where $\xi_i(t)$ are i.i.d. Gaussian variables with zero mean and variance $\sigma^2$. Here, $\zeta_0$ denotes the average rescaled productivity. It will be  convenient to introduce the notation $\zeta$:
\begin{equation}
\zeta := \zeta_0 e^{-\sigma^2}.
\end{equation}
Rare extreme events, namely the ones driving the transition, are the only $\sigma$-dependent features of the dynamics. The precise role of $\zeta$ will become clear in the following sections.

\subsubsection{Reduced dynamics}

Substituting Eq.~\eqref{eq:HP_productivity} into Eqs.~\eqref{e3}---\eqref{e7} and taking the limit $\varepsilon \to 0$ under the consistent scaling of Eq.~\eqref{eq:HP_scaling} leads to a reduced dynamics where the inventories have been fully eliminated (since they are of order $\varepsilon$). While the full derivation is given in Appendix~\ref{app:high-perishability}, we show the main results below. We start by defining the rescaled flow variable
\begin{equation}
\overline{x}_{ij} := z_i(t) X_{ij}(t),
\end{equation}
as well as the average target production across the customers $\mathcal{C}(i)$ of $i$,
\begin{equation}
\overline{y}^\top_i(t) := K^{-1} \sum_{j\in \mathcal{C}(i)} {y^\top_j(t)}.
\end{equation}
Note that both $\bar{x}_{ij}$ and $\overline{y}^\top_i$ are of order one in the limit $\varepsilon \to 0$. The reduced dynamics read
\begin{align}
    y_i(t) &= \min\left[y^\top_i(t),e^{\xi_i(t)-\xi_i(t-1)} \, \min_{j \in \mathcal{S}(i)} \overline{x}_{ij}(t-1)\right], \label{eq:limit_prod} \\
    \overline{x}_{ij}(t) &= \varkappa y^\top_i(t) \min\left[1 ,\displaystyle{\frac{\zeta \,  y_j(t)}{\varkappa \overline{y}^\top_i(t)+{\zeta}c}}\right], \label{eq:limit_x} \\
    y^\top_i(t+1) &= (1-\omega) y^\top_i(t) + \omega \min\left(c+\frac{\varkappa}{\zeta}\overline{y}^\top_i(t),\min_{j \in \mathcal{S}(i)}\overline{x}_{ij}(t)\right). \label{eq:limit_plan}
\end{align}

Each of these equations can be understood in economic terms clearly. Equation~\eqref{eq:limit_prod} is the Leontief production rule: realised output is the smallest between the firm's target and what the weakest supplier can deliver, with a noise factor reflecting the firm's own productivity shock. Equation~\eqref{eq:limit_x} is the rationing rule: when supplier $j$ cannot meet all orders, deliveries to $i$ are scaled down proportionally. Finally, Equation~\eqref{eq:limit_plan} is the target adjustment rule: firms revise their production plan to the minimum between expected demand and physically deliverable supply. 

Critically, all of these equations are expressed in terms of order-one variables only. The inventory state $S_{ij}$ has been eliminated, since it is fully dependent algebraically to the current flows $x_{ij}$. The only persistent dynamics in the high-perishability limit are those of the production targets and the shortages that propagate through the network.

\subsubsection{Phase diagram}

\begin{figure}
    \begin{tikzpicture}
    \node at (0,0) {\includegraphics[width=0.5\linewidth]{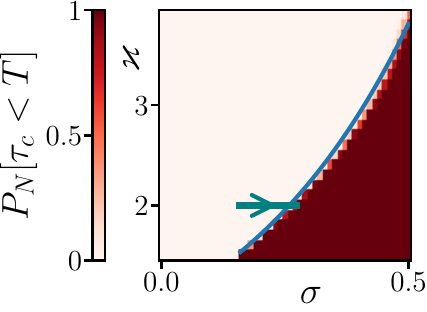}};
    \node[draw,densely dotted, rounded corners, line width=1.1,inner sep=8pt] (ev) at (8,0) {\includegraphics[width=0.42\linewidth]{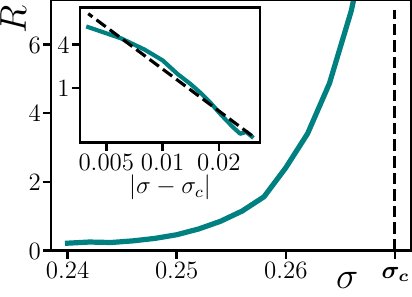}};
    \node[draw, minimum width=2cm, densely dotted, line width=1.5, minimum height=0.8cm,rounded corners] (spot) at (0.95,-0.9) {};
    \draw[densely dotted,line width=1.1] (spot.north east) -- (ev.north west);
    \draw[densely dotted,line width=1.1] (spot.south east) -- (ev.south west);

    \tikzmath{\ss=1.5;}
    \node[scale=\ss] at (-0.6,2.1) {a.};
    \node[scale=\ss] at ($(ev.center)+(2.5,-1.2)$) {b.};
    \end{tikzpicture}
    \caption{High-perishability limit. {\bf a.} Heat map of the empirical crash probability in the $(\sigma, \varkappa)$ plane, for $N=750$ and $T=2000$. Upon increasing $\sigma$, the economy transitions from resilient to fragile beyond a critical $\sigma_c(\varkappa)$ predicted to be $\log(\varkappa)/2\sqrt{\log(K)}$ [see Eq. \eqref{eq:sigmac_main}] and indicated by the thick blue line. For $(\sigma, \varkappa)$ below the boundary line, the economy crashes almost surely. Note that there is a value $\sigma_{\max}$, computed in Eq. \eqref{eq:sigma_max}, beyond which the economy is always unstable, even when inventories are very high. {\bf b.} Excess volatility ratio $R$ (defined in Eq. \eqref{eq:excess_vol}) as a function of $\sigma$ for $\varkappa = 2$, such that $\sigma_c \approx 0.27$ (black dashed line). (Inset) Log-Log plot of excess volatility $R$ as a function of $\sigma - \sigma_c$,  compatible with a divergence $(\sigma_c - \sigma)^{-2}$, the black dashed line indicating a slope of $-1.95$. Parameters: $c=6$, $K=6$, $\zeta_0=30$,  $\omega=0.1$.}
    \label{fig:placeholder}

\end{figure}

Figure~\ref{fig:placeholder}a shows the numerical phase diagram of the system dynamics (\ref{eq:limit_prod}-\ref{eq:limit_plan}) in the $(\sigma, \varkappa)$ plane. The qualitative structure is the same as in the finite-perishability case of Figure~\ref{fig:phase_diagram}a. A sharp boundary $\sigma_c(\varkappa)$ separates a resilient region from a fragile one, terminating at an upper bound $\sigma_{\max}$ above which no level of inventories can prevent a crash. The transition is, again, continuous.

The excess volatility ratio $R$ is, as expected, stronger than for finite perishability (compare Fig. \ref{fig:placeholder}b with Fig. \ref{fig:cumul}b), and appears to diverge faster as the transition line is approached, as $R \propto (\sigma_c - \sigma)^{-2}$. In section \ref{sec:excess_vol}, we calculate the single firm excess volatility (i.e. neglecting correlations between firms) and find that it diverges as $\propto (\sigma_c - \sigma)^{-1}$, suggesting that aggregate volatility has two components: the production fluctuations of each individual firm and their mutual correlations, mediated by propagation along the supply chain. Near the critical line, not only does each individual firm become more volatile, but fluctuations become highly correlated across the network, so aggregate volatility grows faster than the variance of individual shocks.  

The thick blue line in Fig~\ref{fig:placeholder}a is an analytical prediction, $\sigma_c(\varkappa) = \log(\varkappa)/2\sqrt{\log(K)}$, which we derive in the $N\to\infty$ limit in the next section.

\subsection{The critical shock amplitude $\sigma_c$}
\label{sec:HP_critical}

The central analytical result of this section is a closed-form expression for the critical shock amplitude:
\begin{equation}\label{eq:sigmac_main}
    \boxed{\;\sigma_c(\varkappa) = \frac{\log \varkappa}{2 \sqrt{\log K}},\;}
\end{equation}
or, alternatively, the minimum level of inventories needed for the economy to be resilient:
\begin{equation}\label{eq:kappac_main}
    \boxed{\;\varkappa_c(\sigma) = e^{2 \sigma \sqrt{\log K}}.\;}
\end{equation}
In what follows, we will derive this expression and interpret its economic content. The derivation rests on two ideas. The first is that the shock volatility $\sigma$ and the inventory buffer $\varkappa$ enter the dynamics as two competing scales: the buffer can absorb a typical worst-case shortfall up to magnitude $\log(\varkappa)$, while the worst shock among $K$ Gaussian draws scales as $-2\sigma\sqrt{\log K}$. Stability is obtained when the inventory buffer dominates the worst shock, and the critical line indicates where these two terms cross. The second idea is that the dynamics can be written \textit{self-consistently} in terms of two quantities: the average target production $\overline{y}^{\top}(t)$ and the production of the worst supplier, $y_<(t)$. Their coupled evolution determines whether shortages decay or propagate. 

\subsubsection{Steady state of the rescaled dynamics}
\label{sec:HP-critical}

In this section, we identify the steady state of Eqs. (\ref{eq:limit_prod}-\ref{eq:limit_plan}). Suppose for now that production matches its target ($y_i=y^\top_i$, no intermediate input binds production). Equation~\eqref{eq:limit_plan} is therefore reduced to

\begin{equation}\label{eq:steady_state_smalleps}
    y_i(t+1)= (1-\omega) y_i(t) + \omega \left(c+\frac{\varkappa}{\zeta}\overline{ y}_t\right)\;,
\end{equation}
which has the homogeneous fixed point
\begin{equation}
    y^\star =c+\frac{\varkappa}{\zeta} y^\star \to y^\star = \frac{\zeta c}{\zeta - \varkappa}\;. \label{eq:ystar_HP}
\end{equation}

Eq \eqref{eq:ystar_HP} exists when $\zeta>\varkappa$ and coincides with the general steady state in Eq.~\eqref{eq:eqproduction} in the limit $\varepsilon\to 0$ and $K\to\infty$. To assess its stability, we linearise Eq.~\eqref{eq:steady_state_smalleps}, yielding $\Delta y(t+1) = (1 - \omega + \omega\varkappa/\zeta)\, \Delta y(t)$ with a contraction rate $1 - \omega(1 - \varkappa/\zeta) < 1$. Hence, this steady-state always remains stable. 

The requirement that $\zeta>\varkappa$ has important implications, since, from Section~\ref{sec:HP-numerical}, $\zeta=\zeta_0 e^{-\sigma^2}$. It thus reads $\zeta_0>\varkappa e^{\sigma^2}$, showing that for a fixed average productivity $\zeta_0$, the steady state ceases to exist if $\sigma$ is too large --- a first hint that very large shocks can destabilise the economy regardless of inventory buffers.

\subsubsection{Coupled dynamics: average production and worst supplier}

As firm productivities fluctuate, the homogeneous steady state is perturbed. In a finite-size economy with a random regular network structure, the production levels $y_i(t)$ are approximately i.i.d. across firms in the large-$K$ limit, and we can describe their joint behavior by a single distribution. Two quantities of this distribution govern the dynamics:

\begin{itemize}
    \item The \textit{average target production}, $\displaystyle\overline{y}^\top_t = K^{-1}\sum_{j\in \mathcal{C}(i)} y_j^\top(t)$, which becomes $i$-independent in the large-$K$ limit through the law of large numbers;
    \item The \textit{worst supplier's production}, $y_<(t)$, defined as the typical value of $\displaystyle\min_{j\in\mathcal{S}(i)}y_j(t)$ across firms $i$.
\end{itemize}

The first variable captures the bulk of the distribution, while the second captures its left tail, which is what drives the behavior under the Leontief production hypothesis (where production is bottlenecked by the scarcest input).

It is therefore convenient to work with the following dimensionless ratio,

\begin{equation}
r_<(t) := \frac{y_<(t)}{\overline{y}^\top_t},
\end{equation}
which measures by how much the typical worst supplier lags behind the typical firm. 

In a well-defined steady state, all firms have the same production, and so $r_<(t)=1$. When idiosyncratic shocks pull some firms below the average $r_<(t)\lesssim 1$, the question is whether this gap closes back up (and the economy is resilient), or widens (and the economy crashes).

\subsubsection{Dynamics of the average production}

We start with the dynamics of the average target, $\overline{y}^\top_t$. The full derivation, which involves a detailed analysis that depends on whether the bottleneck is a binding constraint, is given in Appendix~\ref{app:HP-derivation}. The result is that two cases should be distinguished. When bottlenecks are binding one finds
\begin{equation}\label{eq:avg_dynamics}
    \overline{y}^\top_{t+1} = \overline{y}^\top_t \left[1 - \omega + 
    \omega\, \varkappa\, r_<(t)\, f\!\left(\overline{y}^\top_t\right)\right],
    \qquad 
    f(y) := \frac{\zeta\, y}{\varkappa\, y + \zeta\, c}.
\end{equation}
The function $f$ is increasing in $y$ with $f(y^\star)=1$, so $f\!\left(\overline{y}^\top_t\right)<1$ whenever the $\overline{y}^\top_t$ runs below its steady-state value. Equation~\eqref{eq:avg_dynamics} has a simple economic interpretation: the average target decays when the worst-supplier ratio $r_<(t)$ is far below $1$. In this case, the fate of the average is directly tied to the fate of the worst supplier.

In the case where bottlenecks are non-binding, one finds a simple reversion to the mean dynamics: 
\begin{equation}\label{eq:avg_dynamics_nonbinding}
    \overline{y}^\top_{t+1} = (1-\omega)\, \overline{y}^\top_t + \omega\left(c + \frac{\varkappa}{\zeta}\, \overline{y}^\top_t\right).
\end{equation}

\paragraph{Economic interpretation of $f(y)$}
The function $f(y)$ has a transparent economic reading: it is the network-averaged rationing rule. Under the proportional rationing scheme~\eqref{e5}, each customer receives a share of supplier $j$'s output that is proportional to how much they ordered. This share, in the homogeneous case we are studying, is exactly $f(\overline{y}^\top_t)$. The shape of $f$ therefore encodes how scarcity is allocated across customers when supply falls short. 


This interpretation allows us to discuss intuitively what the introduction of price flexibility would do. Prices are, by nature, a mechanism for allocating scarce resources. Price-based allocation would reshape $f$ to favour customers with the highest marginal value for the scarcest input, in the spirit of~\citet{Bachmann2024,baqaee_2018_cascading} or~\citet{Elliott2025}. Firms whose production would otherwise be halted by the bottleneck (that is, those with a low $r_<$ \textit{and} a high marginal cost of the missing input) would bid more aggressively and receive a larger share. This would effectively shift $f$ upward and mitigate the effect of cascades. The aggregate effect of price flexibility in this framework is therefore to \textit{redistribute} scarcity away from firms most likely to trigger cascades, raising the critical threshold $\sigma_c$, but not making it disappear. Actually, an extension of the calculations presented in Appendix \ref{app:HP-derivation} shows that aggressively diverting the consumption of rare goods from households to firms would only increase the threshold $\sigma_c$ by a factor $\log \zeta/\log \varkappa > 1$, beyond which instability reappears. When the worst-case shock is large enough, no allocation can prevent some critical firm from running short.

The arguments raised by~\citet{geerolf:hal-04015954} regarding the 2022 European gas crisis suggests that this mitigative role of prices is in fact severely limited at the short-run frequencies that we consider in this analysis: market-clearing prices increases have to be implausible large, something that is prevented by long-term contracts and government prices, implying that, in reality, some of the allocation is in fact done through quantity rationing. This provides a justification for our proportional-rationing benchmark, which captures the regime in which price-mediated reallocation of scarcity is too slow or too constrained to help mitigate cascades. 

\subsubsection{Dynamics of the worst supplier}

We now turn to $y_<(t)$. Each firm's realised production satisfies \eqref{eq:limit_prod}, which we can rewrite using $\Xi_i(t) := \xi_i(t+1) - \xi_i(t)$, a Gaussian variable that now has a variance $2\sigma^2$. 

Following the steps in Appendix~\ref{app:HP-derivation}, we find that realised production for firm $i$ is now
\begin{equation} \label{eq:realized_prod}
    y_i(t+1) = y^\top_i(t+1)\, \min\!\left[1,\; \varkappa\, e^{\Xi_i(t)}\, r_<(t)\, f\!\left(\overline{y}^\top_t\right)\right].
\end{equation}

Taking the minimum over the $K$ suppliers of a given firm and using the approximate homogeneity of target productions, we obtain the following equation
\begin{equation}\label{eq:ymin_evolution_main}
    y_<(t+1) \approx \overline{y}^\top_{t+1}\, 
    \min\!\left(1,\; \varkappa\, \min_{i=1,\ldots,K} e^{\Xi_i(t)}\, r_<(t)\, f\!\left(\overline{y}^\top_t\right)\right).
 \end{equation}

To understand its implications, we now turn to extreme-value theory. The minimum of $K$ i.i.d. Gaussian variables of variance $2\sigma^2$ behaves, for large $K$, as 
\begin{equation}
    \min_{i = 1,\ldots,K} \Xi_i \approx -2\sigma\sqrt{\log K},
\end{equation}
so the typical worst shock among the $K$ suppliers contributes a factor $e^{-2\sigma\sqrt{\log K}}$.\footnote{A more refined reasoning, which includes fluctuations around this typical value, can be made for the volatility calculation in Appendix F and is given there.} Substituting and dividing by $\overline{y}^\top_t$ gives the linearised update rule for the ratio,
\begin{equation}\label{eq:r_dynamics}
    r_<(t+1) = \gamma\, r_<(t)\, f\!\left(\overline{y}^\top_t\right), 
    \qquad \gamma := \varkappa\, e^{-2\sigma\sqrt{\log K}}=e^{-2\sqrt{\log K} (\sigma-\sigma_c)},
\end{equation}
where we use that $\sigma_c := \log \varkappa/2 \sqrt{\log K}$.\footnote{The i.i.d. assumption for $\xi_i(t)$ is not essential. If the shocks have a correlation $\rho$, the variable $\Xi_i(t)$ is Gaussian with variance $2(1-\rho)\sigma^2$, and all results hold by replacing $\sigma\to\sigma(1-\rho)$. Persistence therefore \textit{stabilizes} the economy: since inventories depend directly on flows, only productivity \textit{fluctuations} affect production, and persistent shocks ($\rho\to 1$) do not trigger cascades. }

This  is one of our central results and contains everything we need to understand the transition. It states that the worst-supplier ratio $r_<$ evolves multiplicatively with the rate $\gamma$, scaled by the factor $f(\overline{y}^\top)<1$. The single dimensionless parameter $\gamma$ governs whether the tail of the distribution (driven by $y_<$) catches up with the core (given by $\overline{y}^\top_t$) or falls behind. 
Equations~\eqref{eq:avg_dynamics} and \eqref{eq:r_dynamics} form a closed two-dimensional system. The qualitative behavior is determined entirely by the value of $\gamma$:

\begin{itemize}
    \item If $\gamma>1$ (i.e. $\sigma<\sigma_c$) then the $\min$ in the update-rule~\eqref{eq:ymin_evolution_main} for $y_<$ saturates at $1$. The worst-supplier ratio is reset to $r_<(t+1)=1$ each period, and Eq.~\eqref{eq:avg_dynamics} keeps the average at $y^\star$. The economy is therefore resilient. 
    \item If instead $\gamma<1$ (i.e. $\sigma>\sigma_c$), then the worst-supplier ratio contracts geometrically as $r_<(t)\sim \gamma^t\to 0$, and once the bracketed factor in \eqref{eq:avg_dynamics} dips below one, the average $\overline{y}^\top_t$ begins to decay too. The economy crashes, and the steady state is not resilient.
\end{itemize}
Thus, $\gamma>1$ is equivalent to $\sigma<\sigma_c$ and vice-versa.

\subsection{Comparative statics and consequences}

Equation~\eqref{eq:sigmac_main} delivers several economically meaningful consequences.

\paragraph{Decreasing in $K$.} Counterintuitively, $\sigma_c$ \textit{decreases} in the number of inputs each firm requires. However, $K$ is the number of \textit{bottlenecks}. Under Leontief complementarity, more inputs means more opportunities to encounter a critical shortfall: the worst case shock across $K$ Gaussian draws scales as $-2\sigma\sqrt{\log K}$ and erodes the buffer provided by inventories. Read literally, this seems to suggest that firms with many suppliers are systematically more fragile --- which is clearly not what is observed in practice.

The resolution is that the Leontief assumption is more stringent than realistic production technologies. In practice, most inputs are at least partially substitutable. A manufacturer short on one steel grade can often use a related one. What our analysis isolates, instead, is the fragility caused specifically by \textit{non-substitutable} inputs, the ones whose absence genuinely halts production: energy in most manufacturing, specific electronic components, raw materials with no near term replacement (for instance, lithium for batteries, or gallium for semi-conductors). The parameter $K$ should therefore be read as the number of such \textit{critical bottleneck inputs}, not the total number of suppliers a firm has on its books.

This reframing reverses the apparent paradox. A well-diversified firm is one whose production technology has been designed to avoid such single-supplier dependencies wherever substitution is feasible: it has \textit{fewer} bottleneck inputs, hence a smaller $K$ and a larger $\sigma_c$. The model is silent about substitutable inputs by construction, but identifies the inputs that matter for fragility as those without close substitutes. 

Section~\ref{ref:hp-diversification} takes a complementary perspective to this: holding the number of bottlenecks $K$ fixed, we ask what happens also when firms can choose between $M$ alternative suppliers for each bottleneck input. We will see that supplier diversification along this margin shifts $\sigma_c$ \textit{upward} (or, respectively, $\varkappa_c$ \textit{downward}) as $\sqrt{M}$, and that for large enough $M$ the cascade mechanism can disappear entirely.

\paragraph{Increasing in $\varkappa$, but only logarithmically.} Larger inventory buffers raise the critical shock amplitude. But the dependence $\sigma_c \propto \log\varkappa$ is weak. Doubling inventories does not double resilience. This is consistent with the observation that just-in-time supply chains have been profitable for decades despite producing the occasional dramatic disruption (such as during the Covid-19 pandemic, the Japan 2011 Tohoku earthquake or the 2026 closure of the strait of Hormuz): marginal increases in buffer stocks deliver diminishing returns in stability, and firms  perceive most additional inventory as wasteful in normal times. 

\paragraph{Diverging crash time near the boundary.} When $\sigma$ is just above $\sigma_c$, the worst supplier ratio decays as $r_<(t)\approx \gamma^t$ with $\gamma<1$, even though the average production remains close to $y^\star$ until the bracketed factor in Eq.~\eqref{eq:avg_dynamics} dips below $1$. This crossover defines a typical crash time,

\begin{equation}\label{eq:crash_time_main}
    t_\times \approx \frac{\log\varkappa}{2\sigma\sqrt{\log K} - \log\varkappa} = \frac{\sigma_c}{\sigma - \sigma_c}.
\end{equation}

As $\sigma\searrow \sigma_c$ from above, $t_\times\to\infty$: just past the boundary, the economy can appear stable for arbitrarily long periods while remaining structurally fragile. This has an important practical implication: an economy operating just inside the unstable region looks indistinguishable from a resilient one over any finite observation window, and crashes appear to arrive without warning. The model thus provides a mechanism for ``black-swan''-like supply chain collapses that follow extended periods of apparent stability.

\paragraph{Upper bound $\sigma_{\max}$.} The stability condition $\sigma<\sigma_c(\varkappa)$ must be combined with the steady state existence condition $\zeta_0>\varkappa e^{\sigma^2}$ derived in Section~\ref{sec:HP_critical}. Together, they place $\log\varkappa$ in the window $[2\sigma\sqrt{\log K}, \, \log\zeta_0 - \sigma^2]$. As $\sigma$ grows, the lower bound rises (stability is more and more demanding), while the upper bound falls (steady state existence becomes more demanding). The window closes entirely at 
\begin{equation}\label{eq:sigma_max}
    \sigma_{\max} = \sqrt{\log\!\big(K \zeta_0\big)} - \sqrt{\log K},
\end{equation}
above which no level of inventory restores stability. (We recall that $\zeta_0 > 1$ for equilibrium to exist in the first place). This is the upper bound visible in the numerical phase diagrams (Figs.~\ref{fig:phase_diagram} and \ref{fig:placeholder}-a). The existence of $\sigma_{\max}$ also has a clear economic interpretation: very large shocks destabilise the economy not by overwhelming inventories per se, but by violating the Hawkins-Simon productivity-vs-buffer balance\footnote{Note that the standard Hawkins-Simon condition would rather frame it as a productivity vs. demand balance, but buffers drive demand in this model.} that makes a steady state possible.

\subsection{Stabilizing an unstable economy}
\label{ref:hp-diversification}

The previous subsection raised questions that remains unresolved. The parameter $K$ corresponds to the number of \textit{bottleneck} inputs, for which the firm has no substitutes and whose absence halts production entirely. We found that bottlenecks erode resilience, as corresponds to intuition. Here, we ask another complementary question: what happens if the firm has access to $M$ alternative suppliers for each bottleneck input? This follows the ideas outlined in \citet{barrot_specificity_2016, taschereau2025cascades}.

Let us therefore write that for each of the $K$ input bottlenecks, firms can choose between $M$ different suppliers. More precisely, we posit that at each time step firms choose the supplier $\alpha= 1, \dots, M$ with the largest production of each good $i=1, \dots, K$. The previous analytical analysis remains valid provided we now interpret $y_<$ as the minimum over $K$ goods of the maximum over $M$ possible suppliers of the current production. The evolution of $y_<(t)$ then becomes (compare with Eq. \eqref{eq:ymin_evolution_main}):
\begin{align}
y_<(t+1) \approx \overline{y}^\top_{t+1} \, \min\left(1,\varkappa \min_{i=1,\dots,K} \left[ \max_{\alpha=1,\dots,M} e^{\Xi_{i\alpha}(t)}\right]\frac{y_<(t)}{\overline{y}^\top_t}f\left(\overline{y}^\top_t\right)\right) \;.
\end{align}
Interestingly, one can show, using standard results on the distribution of extrema of independent random variables, that when $M > \log K/\log 2$ the most probable value of $\Xi_m := \min_{i=1,\dots,K} \left[\max_{\alpha=1,\dots,M} {\Xi_{i\alpha}(t)}\right]$ becomes {\it positive} for i.i.d. Gaussian random variables, meaning that the probability that $\Xi_m$ is negative and that $y_<$ decreases with time becomes extremely small. In other words, when firms have a large number of alternative suppliers, and productivity shocks are independent, crises can be avoided thanks to immediate ``rewiring''. Of course, the assumptions that shocks are independent and that rewiring can take place at each time step are not realistic and in practice crises can still appear for large, global productivity shocks. 

When  $M \ll \log K/\log 2$, on the other hand, one can show that $\Xi_m \approx -2 \sigma \sqrt{(\log K)/M}$, leading to a $\sqrt{M}$ increase of the critical value of the amplitude of productivity shocks [compared to Eq. \eqref{eq:sigmac_main}]:
\begin{equation}
     \sigma_c(M) = \sqrt{M} \frac{\log \varkappa}{2 \sqrt{\log K}}.
\end{equation}
Therefore, as expected intuitively, supplier diversification makes the firm network more resilient to shocks \citep{barrot_specificity_2016, taschereau2025cascades}, until the instability disappears when $M$ is large enough. For example,  when $K=10$, $M$ should be larger than $\log K/\log 2 \approx  3.32$, i.e. three alternative suppliers for a given good are needed for the economy to be fully resilient. 

However, note that rewiring must be fast enough to prevent the propagation of production failures along the supply chain. More precisely, the time needed to switch supplier must be short compared to the crash time $t_\times$ given in Eq. \eqref{eq:crash_time_main}. A numerical simulation of such a dynamical rewiring would be needed to clarify the precise behavior of the model in this case. For more recent work on the role of rewiring in production networks, see e.g.\ \citet{acemoglu_azar_2020}, \cite{colon2022radical} and \cite{taschereau2025cascades}.

Finally, the closed-form expression given in Eq.~\eqref{eq:sigmac_main} reproduces the qualitative shape of the numerical phase boundary shown in Fig.~\ref{fig:placeholder}a (blue line) without any fitting parameters. The next subsection examines what happens as the boundary is approached from the resilient side, where the economy does not crash but exhibits divergent excess volatility $R$.

\section{Excess volatility, Self-Organized Criticality \& Resilience}
\label{sec:excess_vol}

\subsection{From individual firms to system wide shocks}

When $\sigma <\sigma_c$ (or equivalently $\varkappa > \varkappa_c$), the economy does not crash, but the volatility of aggregate productivity diverges as the critical line is approached. We saw this numerically in Fig.~\ref{fig:placeholder}-b, where the excess volatility ratio $R$ defined in Eq.~\eqref{eq:excess_vol} appeared to diverge as $R\sim (\sigma_c-\sigma)^{-2}$. We now derive the single-firm volatility analytically and show that aggregate volatility diverges \textit{faster} than single-firm fluctuations. We interpret this as a signature that fluctuations become increasingly correlated across the network as the boundary is approached.

\subsubsection{Dispersion of the worst supplier}
\label{subsubsec:dispersion_supplier}

Close to the transition, the linearised dynamics of the worst-supplier ratio (written in Eq.~\eqref{eq:r_dynamics}) can be modelled as a random walk for $\phi(t)=-\log r_<(t)$. Defining 
\begin{equation} \label{eq:eta_def}
    \eta:=\log\varkappa - 2\sigma\sqrt{\log K} \equiv 2 \sqrt{\log K} \, (\sigma_c - \sigma)
\end{equation}
as the distance to the transition (that is, with $\eta>0$ corresponding to the resilient regime $\sigma<\sigma_c$), and using a refined version of the extreme-value asymptotics that retains the fluctuations around the typical worst shock,
\begin{equation}\label{eq:gumbel_refined}
    \min_{i=1,\ldots,K} \Xi_i \approx -2\sigma\sqrt{\log K}\left(1 + \frac{w}{2\log K}\right),
\end{equation}
where $w$ is a Gumbel-distributed random variable of mean $\bar w = 0.577\ldots$ (the Euler-Mascheroni constant) and variance $\pi^2/6$, one obtains the dynamics 
\begin{equation}\label{eq:phi_dyn}
    \phi(t+1) = \left[\phi(t) - \eta + \frac{\sigma}{\sqrt{\log K}}(w(t) - \bar w)\right]^+.
\end{equation}

The positive part $[\cdot]^+$ ensures $\phi \geq 0$ (or  equivalently $r_< \leq 1$): the worst supplier cannot exceed the average. Equation~\eqref{eq:phi_dyn} describes a random walk with negative drift $-\eta$ and Gumbel shocks of size $\sigma/\sqrt{\log K}$, reflected at the origin. The full derivation of \eqref{eq:phi_dyn} is given in Appendix~\ref{app:rare-events}.

\subsubsection{Single-firm divergence} 

For $\eta>0$, Eq.~\eqref{eq:phi_dyn} has a stationary distribution that combines a Dirac-delta at the origin (a point mass indicating that $r_<(t)=1$ and the worst supplier matches the average most of the time) with an exponential tail for $\phi>0$, corresponding to occasional excursions where the worst supplier lags significantly behind:
\begin{equation}\label{eq:phi_distribution}
    P(\phi) = Z\,\delta(\phi) + (1-Z)\, B\, e^{-B\phi},
\end{equation}
with $B = (12\log K / \pi^2)\, \eta/\sigma^2$ and $Z$ a coefficient of  order unity. The variance of $\phi$ is therefore
\begin{equation}
    \mathbb{V}[\phi] = \frac{1-Z}{B^2} \propto \eta^{-2}.
\end{equation}

Since we have that $\eta\propto (\sigma_c-\sigma)$ close to the transition, this gives a single-firm divergence,
\begin{equation}\label{eq:single_firm_divergence}
    \mathbb{V}[\phi] \propto (\sigma_c - \sigma)^{-2}, \qquad 
    \text{equivalently } \sqrt{\mathbb{V}[\phi]} \propto (\sigma_c - \sigma)^{-1}.
\end{equation}

The log productivity gap between the worst-performing firm and the average firm grows as $(\sigma_c-\sigma)^{-1}$ as we approach the transition. This is a signature of critical fluctuations in the firm-level output distribution.

\subsubsection{Aggregate volatility and the comovement cluster}

Equation \eqref{eq:single_firm_divergence} accounts only for the fluctuations of individual firms in isolation. The aggregate excess volatility ratio $R$ measured numerically in Fig~\ref{fig:placeholder}-b diverges as $(\sigma_c-\sigma)^{-2}$, which is one power of the $\sigma$-gap faster than the single-firm dispersion. The discrepancy reflects the growing role of inter-firm covariance.

To see why, decompose the variance of aggregate output into a single-firm part and a correlation part:
\begin{equation}
    \mathbb{V}\!\left[\sum_i y_i\right] = \sum_i \mathbb{V}[y_i] + \sum_{i \neq j} \mathrm{Cov}(y_i, y_j).
\end{equation}

Far from the transition, shocks are absorbed locally and covariances are small. Aggregate volatility scales as $\sqrt{N}$ times single-firm volatility, and $R$ remains close to $1$. As the transition is approached, however, shortages propagate further through the network before being absorbed. A perturbation at firm $j$ now affects the production of firms several steps downstream. The covariance term grows faster than the single-firm-variance term, and aggregate volatility diverges accordingly. 


The economic content of this result is that the number of firms whose outputs comove with any given firm -- the \textit{comovement cluster}, or in statistical physics-terminology the 
\textit{correlation volume}
 -- grows without bound near the critical line. This is a generic feature of continuous transitions and provides the mechanism by which idiosyncratic shocks generate macroeconomic fluctuations: near the boundary, shocks no longer ``average out'' across firms even when the network has many nodes. See~\cite{moran2019may} for a more detailed discussion of this mechanism in the context of production networks.
 
\subsection{Endogenizing $\varkappa$: a Minsky-like path to self-organised criticality}
\label{sec:soc_endogenous}

In this subsection, we argue that competitive pressure on inventory costs can drive the economy to operate close to the fragility boundary. The mechanism is a network externality. When a firm trims its own buffer it lowers its own holding cost, but it also raises the chance that its own customers are deprived of inputs during a shortage, and hence the crash risk borne by the rest of the network. Because this spillover is not priced, individually optimal inventories are collectively too lean --- the inventory analogue of the underinvestment in supply-relationship reliability of~\cite{Elliott2022}. We now make this precise and show that the decentralised economy may either settle at a stable buffer  $\varkappa^\star > \varkappa_c$ or be driven to the critical point itself.

We proceed in three steps. First, we write the cost a firm incurs from production drops when $\varkappa$ is small relative to the shock amplitude. 
Second, we impose firms to equate the cost of holding their inventories with their expected losses. 
We show that enforcing the latter balance leads to two regimes. Depending on the cost of holding, it is possible that an optimal solution $\varkappa^\star>\varkappa$ exists and each firm will choose an inventory level that makes the economy stable; the other regime, instead, causes the economy to settle right at the critical boundary $\varkappa=\varkappa_c$, leading therefore to a self-organised critical state.

From the point of view of firms, holding goods in excess of immediate needs entails several types of costs: transportation, storage, and perishability costs, since a fraction of unused goods must be replenished at each time step. For simplicity, we assume that these costs sum to a quantity $C \varkappa$, proportional to the level of precautionary inventories $\varkappa$.

It is rational for firms to incur such costs insofar as excess inventories allow them to maintain production when some input goods cannot be procured. Indeed, production shortfalls also generate costs (e.g., fixed costs and wages) that are not offset by sales.

What, then, is a firm's expected production shortfall during a period when the economy operates normally? We place ourselves in a scenario where firms believes to be subject to an idiosyncratic shock which is not affecting the entire chain. 
In this setting, each firm maintains its own target production $y_i^\top$ unchanged and equal to $y^\star$, thereby leaving the averaged planned production $\overline{y}^\top$ unaffected. 
This is justified as long as the update term in \eqref{eq:avg_dynamics} remains close to $0$, which holds when the update time $1/\omega$ remains higher than the crash time $t_\times$, corresponding to $\omega \ll \eta$. 
In this regime, even if $r_{<}$ has dropped sufficiently below $1$, the target production remains unaffected because $\omega$ is small enough.
To model a shortfall, we assume that $r_{<}$ follows a path of length $\tau_1$ such that $r_{<}(0)=r_{<}(\tau_1)=1$ while $r_{<}(t)< 1$ for all $0<t<\tau_1$: at least one of the firms' suppliers is ailing for a period $\tau_1$.
In this specific scenario, the production of the firm is given by Eq.~\eqref{eq:realized_prod} as 
\begin{align}
    \label{eq:realized_prod_SOC}
    y_i(t+1) = y^\star\, \min\!\left[1,\; \varkappa\, e^{\Xi_i(t)}\, r_<(t) \right].
\end{align}
From Eq.~\eqref{eq:realized_prod_SOC}, we observe that when $\varkappa e^{\Xi_i(t)} r_<(t) \lesssim 1$, production must contract and fixed costs arise.
The total anticipated cost associated with a drop in procurement can be written as
\begin{equation}\label{eq:cost_drop}
    P \sum_{t=1}^{\tau_1}  \left(y^{\star} - y^{\star} \min\!\left[1,\; \varkappa\, e^{\Xi_i(t)}\, r_<(t) \right]\right) = y^\star P \sum_{t=1}^{\tau_1}  \left(1 - e^{-\left[\phi(t) - \log \varkappa - \sigma^2\right]^+}\right)
\end{equation}
where, $\phi := - \log r_< > 0$ evolves according to Eq.~\eqref{eq:phi_dyn}, and the random term $e^{\Xi_i}$ has been replaced by its mean. 
The price of goods $P$ measures the magnitude of the fixed costs that are compensated, in equilibrium, by sales.

Taking the continuous-time limit of Eq.~\eqref{eq:phi_dyn} amounts to modeling $\phi_t$ as a Brownian motion with negative drift:
\begin{equation}
    {\rm d}\phi_t = - \eta \, {\rm d}t + \Gamma \, {\rm d}B_t,
\end{equation}
where $\Gamma := {\pi \sigma}/{\sqrt{6 \log K}}$. We also assume that the shock amplitude $\sigma$ is fixed and known to firms. From Eq.~\eqref{eq:eta_def}, one has $\eta = \log (\varkappa/\varkappa_c)$, which vanishes as $\varkappa \searrow \varkappa_c$.  

In this continuous time limit, the expectation of Eq.~\eqref{eq:cost_drop} can be computed explicitly and reads (see Appendix \ref{app:shortfall}):
\begin{equation} \label{eq:average_cost}
     y^\star P \, \frac{1}{\eta} \, \frac{\Gamma^2}{2\eta + \Gamma^2} \, e^{-2\eta (\log \varkappa + \sigma^2)/\Gamma^2},
\end{equation}
with an average duration of these excursions equal to the factor $1/\eta$ in this formula. Hence, the average production loss over a long time $T$ is given by $T \eta$ multiplied by Eq. \eqref{eq:average_cost}. The last exponential term in this equation gives the probability that a ``crisis'' (i.e. a production shortfall) occurs. 

It is then reasonable to assume that firms will agree to pay an extra inventory cost $C\varkappa T$ as an insurance premium covering the expected losses due to adverse conditions. Hence, a central planner should ensure that $\varkappa$ settles at a value such that
\begin{equation}\label{eq:eq_costs}
   C \varkappa = {y^\star P} \, \frac{\Gamma^2}{(2 \log (\varkappa/\varkappa_c) + \Gamma^2)} \, e^{-2 \log (\varkappa/\varkappa_c) (\log \varkappa + \sigma^2)/\Gamma^2}.
\end{equation}
Note that for $\varkappa > \varkappa_c$, the right-hand side of Eq.~\eqref{eq:eq_costs} is a monotonically decreasing function, bounded from above by ${y^\star P}$. Therefore, when the minimal inventory costs $C \varkappa_c$ required for economic stability are smaller than the one-period production costs $y^\star P$, Eq.~\eqref{eq:eq_costs} has a unique solution with $\varkappa^\star > \varkappa_c$, where the economy is stable, and that the central planner should strive to enforce.  

Can firms individually learn how to coordinate around this equilibrium value? The problem is that as soon as $\varkappa^\star - \varkappa_c \gtrsim \sigma/(\log K)^{3/2}$, crises become extremely rare [see the last term in Eq. \eqref{eq:average_cost}]. This means that over a rather long period, such crises do not occur and complacency {\it à la Minsky} leads to free-riding behavior \citep{minsky2008stabilizing}: firms are tempted to reduce their inventory costs and move towards a ``just-in-time'' model, driving the system closer to criticality -- until the probability of crises suddenly increases.\footnote{This scenario is very similar to free-riding behavior in the context of vaccination: individuals dodge vaccines (i.e. reduce $\varkappa$) relying on collective immunity (i.e. low probability of crises).} In such a system, one expects to observe the analog of Minsky cycles -- i.e. periods of stability interrupted by self-generated periods of fragility, as argued in e.g. \cite{bouchaud2024self}.  

In the unrealistic case where $C \varkappa_c > y^\star P$, Eq.~\eqref{eq:eq_costs} has no solution -- the balance between insurance premium and expected losses cannot be achieved. In this case, the system is expected to settle close to the critical point $\varkappa \approx \varkappa_c$, where precautionary inventories are ``just enough'' to avoid collapse. 

\section{Conclusion}

In this paper, we developed a dynamical model of production networks in which firms use Leontief technologies, manage perishable inventories, and face idiosyncratic productivity shocks. Our focus has been on short-run, quantity-constrained dynamics in supply chains, rather than on frictionless, price-clearing equilibria. Within this framework, we have shown that the interaction of network structure, inventory policies, and stochastic productivity leads to a robust resilience–fragility trade-off and a well-defined critical line in parameter space.

From an economic perspective, our main contribution is to provide a dynamically explicit, firm-level mechanism through which purely idiosyncratic shocks can generate persistent, system-wide  volatility and occasional large collapses in output, in the absence of explicit aggregate shocks. Firms face a trade-off between efficiency and resilience: low inventories minimize storage costs but leave the system vulnerable to cascading shortages. As firms economize on precautionary stocks, the economy is endogenously driven closer to a critical region in which small, local shocks can propagate through the input–output network and have disproportionately large aggregate effects. This provides a dynamic, out-of-equilibrium counterpart to the ``small shocks, large cycles'' puzzle highlighted by \citet{Bernanke1994}, and complements the granularity and equilibrium network literatures by emphasizing the role of disequilibrium quantity constraints and inventory dynamics.

On the theory side, our analysis reveals four main results:

\begin{enumerate}
    \item  {\it Existence of a resilient–to-fragile transition\/}. For a given network connectivity and average productivity, there is a critical line in the plane spanned by the volatility of idiosyncratic shocks ($\sigma$) and the size of precautionary inventories ($\kappa$). Above this line, the economy is resilient: aggregate output fluctuates around a well-defined steady state, and although shortages occur, they remain localised. Below the line, the economy almost surely experiences a system-wide  collapse in finite time. We show that this transition is continuous -- there is no discrete jump in aggregate behavior -- and the typical time to collapse diverges as the critical line is approached from below.
    \item {\it Excess volatility near the critical line\/}. Even in the resilient region, aggregate production volatility is substantially amplified by network effects. We define an ``excess volatility'' ratio $R$ that compares fluctuations in aggregate output to fluctuations in productivity. The latter increases sharply as $\sigma$ approaches its critical value $\sigma_c(\kappa)$ [or when $\kappa$ approaches $\kappa_c(\sigma)$], and our results indicate that it diverges at the transition. Thus, the model provides a concrete mechanism through which micro shocks, which would be innocuous in a frictionless, diversified economy, actually generate large macro volatility once realistic inventories and adjustment dynamics are accounted for.
    \item {\it Analytical characterization and role of diversification\/}. In a high-perishability, large-economy limit, we obtain closed-form expressions for the equilibrium output level, the critical volatility $\sigma_c(\kappa)$, and the time scale for the unfolding of crises. We also show that when firms can rapidly switch among multiple potential suppliers for each bottleneck input, the critical volatility is shifted upwards and the system becomes more resilient. In the extreme case where supplier diversification is sufficiently large and rewiring is effectively instantaneous, the critical line disappears and the economy remains stable for all $\sigma$. This highlights the role of both inventories and ex-ante redundancy in supplier relationships as structural determinants of macroeconomic resilience.
    \item {\it Endogenous inventory choice and the fragility trap}. When firms rationally optimize their inventory holdings by balancing storage costs against expected losses from production shortfalls, competitive pressure can drive the economy toward the critical boundary. We show that when inventory costs are sufficiently low relative to production losses, firms optimally choose $\varkappa^\star > \varkappa_c$, sustaining a stable economy. However, the rarity of crises far from criticality induces Minsky-like complacency: firms free-ride on collective stability by reducing inventories, gradually pushing the system back toward the critical point $\varkappa_c$. This mechanism reveals how market-driven behavior can endogenously generate the cyclical alternation between periods of false stability and sudden fragility that characterizes real economies, providing a concrete pathway to self-organized criticality through decentralized decision-making.

\end{enumerate}

These findings speak directly to current debates on the macroeconomic consequences of supply-chain disruptions, such as those observed during the Covid-19 pandemic, the 2022 European gas crisis, and various trade and geopolitical shocks. Traditional models of production networks in macroeconomics, based on Cobb--Douglas or CES technologies and comparative statics \citep[e.g.][]{acemoglu2012network, baqaee_2018_cascading}, have been instrumental in quantifying how shocks propagate across sectors {\it in equilibrium}. However, by construction they abstract from the short-run, quantity-rationed dynamics that arise when inventories, delivery lags, and myopic production decisions interact. Our model, while highly stylised, is designed precisely to capture those dynamics, in the spirit of ARIO-type disaster models and the business-operations literature on supply chains, but in a framework simple enough to allow for analytical progress.

A key message is therefore that disequilibrium quantities and real rigidities in supply chains can be a first-order source of macroeconomic volatility, even when prices are free to adjust in the long run. In the short run, firms face hard constraints: if a critical input is not available, production cannot be instantly substituted, and prices cannot conjure up missing physical inventories. In this sense, our model is closer to the older disequilibrium macro tradition \citep{barro1971general, benassy2014macroeconomics} than to contemporary New Keynesian models. In short, we argue that macroeconomic fluctuations and crises can be driven not only by large aggregate shocks or a few ``mega‑firms,'' but also by the interaction of production networks with short‑run quantity constraints (capacity, inventory, supply shortages) and imperfect adjustment rules. These dynamics are naturally analysed in a non‑Walrasian disequilibrium framework, but extended to modern production networks and enriched with tools from the theory of complex systems and critical phenomena. Of course, a complete theory would have to account for longer time scales as well, and describe if and how the economy is able to eventually revert to equilibrium \cite{Dessertaine2022}.

From a policy perspective, our results suggest several implications. First, there is no single ``optimal” inventory level from a social standpoint once one recognizes that private incentives push firms toward lean, just-in-time operations. Cost minimization tends to erode buffers and move the system closer to the fragile phase, where volatility and the risk of large crises increase sharply. Second, policies that encourage moderate precautionary inventories or diversification of suppliers can have highly non-linear benefits: beyond a certain threshold, they can shift the economy from a crisis-prone regime to a resilient one. Third, public interventions that act on inventories of strategic goods (energy, key intermediates) or that facilitate fast rewiring of supply chains may be particularly powerful near the critical line, where marginal changes in buffers have outsized effects on systemic stability.

Our analysis also opens several avenues for future work. One natural extension is to reintroduce prices and financial constraints, along the lines of \citet{Dessertaine2022}, into our inventory-based dynamics. This would allow us to study how quantity constraints, price adjustments, and credit conditions jointly shape the resilience–fragility boundary. Another important direction is to incorporate heterogeneity in technologies, inventory policies, and firm sizes to connect more directly with granular and firm-level empirical work. A third one is along the lines of experimental microeconomics, where one can bring many of the operating principles to human subjects under controlled laboratory settings, and get rid of synchronous updates of firms' stocks and inventories on a schedule as it is done here. 
Finally, the model's parameters map onto observables, which opens the door to empirical calibration and validation, along the lines of~\cite{hamid2025differentiablemodelsupplychainshocks}. The buffer $\kappa$ corresponds to a firm's input inventory measured in periods of cover, as reported e.g. as the inventory-to-sales ratio in the U.S. Census/BEA series. The shock amplitude $\sigma$ can correspond to the dispersion of firm-level productivity, which could be measured in firm-level panels. Lastly, $K$ is the number of critical, non-substitutable inputs. This means that the predictions of our model (such as the link between volatility and inventory level) can be taken to the data.

Finally, there could be scope for empirical calibration and validation: our model generates clear comparative statics --- relating volatility, crash frequency, and supply --- chain disruptions to observables such as input perishability, inventory-to-sales ratios, and supplier diversification—that can be taken to data.

Taken together, these results underscore that production networks are not just passive amplifiers of shocks in a static equilibrium but complex dynamical systems whose resilience depends sensitively on operational choices --- inventory management, supplier strategies --- and whose endogenous tendency toward efficiency can make them structurally fragile \citep{colon2020fragmentation}. Recognizing and quantifying this trade-off is essential if we are to understand, and eventually design, macroeconomic systems that are both efficient and robust \citep{hynes2022systemic, Moran2025-xm, bouchaud2024self}.

\subsection*{Acknowledgements} We thank S. Chelly, C. Colon, X. Gabaix, S. Gualdi, S. Hamid, D. Luongo, A. Mandel, L. Mungo, M. Nirei, F. Pijpers, J. Scheinkman, S. Towers, U. Weitzel and F. Zamponi for useful discussions on these topics. We also thank C. Colon, S. Hallegate, and M. Taschereau-Dumouchel for constructive comments on the manuscript.  

\bibliographystyle{apalike}

\newpage

\section*{List of symbols and definitions}

\begin{center}
\setlength{\LTcapwidth}{0.9\textwidth} 
\begin{longtable}{ll}
\caption{Main symbols and parameters used in the model}
\label{tab:symbols}\\
\toprule
\textbf{Symbol} & \textbf{Definition / Interpretation} \\
\midrule
\endfirsthead

\toprule
\textbf{Symbol} & \textbf{Definition / Interpretation} \\
\midrule
\endhead

\midrule
\multicolumn{2}{r}{\textit{Continued on next page}}\\
\bottomrule
\endfoot

\bottomrule
\endlastfoot

$N$ & Number of firms (and goods) in the economy \\
$i,j$ & Firm (and good) indices \\
${\cal S}(i)$ & Set of suppliers of firm $i$ \\
${\cal C}(i)$ & Set of customers of firm $i$ \\
$K$ & Number of suppliers (and customers) per firm (network degree) \\
\midrule
$y_i(t)$ & Actual production of firm $i$ at time $t$ \\
$y_i^\top(t)$ & Target (planned) production of firm $i$ at time $t$ \\
$g_i(t)$ & Inventory of finished good $i$ held by firm $i$ at time $t$ \\
$S_{ij}(t)$ & Inventory at firm $i$ of input good $j$ at time $t$ \\
$O_{ij}(t)$ & Order placed by firm $i$ to supplier $j$ at time $t$ \\
$M_{ij}(t)$ & Share of good $j$ allocated to firm $i$ at time $t$ (before truncation) \\
$X_{ij}(t)$ & Actual quantity of good $j$ delivered from $j$ to $i$ at time $t$ \\
$c_j$ & Final consumption demand for good $j$ (often $c_j \equiv c$) \\
$C_j(t)$ & Quantity of good $j$ sold to final consumers at time $t$ \\
\midrule
$z_i(t)$ & Productivity of firm $i$ at time $t$ \\
$z$ & Average productivity level (without shocks) \\
$\xi_i(t)$ & Idiosyncratic productivity shock of firm $i$ at time $t$ \\
$\sigma$ & Standard deviation of idiosyncratic productivity shocks $\xi_i$ \\
$L_0$ & Available labour input per firm (assumed non-binding in baseline) \\
\midrule
$\psi_j$ & Perishability rate of input good $j$ (per period) \\
$\psi$ & Common perishability rate of all goods (we set $\psi_j \equiv \psi$) \\
$\kappa$ & Inventory buffer parameter (fractional safety stock of one-period use) \\
$\omega$ & Adjustment (learning) rate of production targets $y_i^\top$ \\
\midrule
$y^\star$ & Homogeneous equilibrium production level (shock-free case) \\
$S^\star, X^\star, O^\star, g^\star$ & Equilibrium values of $S_{ij}, X_{ij}, O_{ij}, g_i$ \\
$\Delta y, \Delta a, \Delta S, \dots$ & Homogeneous perturbations around the stationary state \\
${\cal L}^+, {\cal L}^-$ & Linear evolution matrices in the demand- and supply-limited cones \\
$\kappa_{\min}$ & Minimal inventory level for which equilibrium exists, $1/(1-\psi)$ \\
$\kappa_c^\star$ & Upper bound on $\kappa$ for equilibrium existence, $(z-K)/(K\psi)$ \\
$\kappa_c^+, \kappa_c^-$ & Critical inventory levels for instability of ${\cal L}^+$ and ${\cal L}^-$ \\
\midrule
$\overline{x}_{ij}$ & Rescaled inventory/flow variable (high-perishability limit) \\
$\varepsilon$ & Small parameter, $\psi = 1 - \varepsilon$ in high-perishability limit \\
$\varkappa$ & Rescaled buffer parameter, $\kappa = \varkappa/\varepsilon$ \\
$\zeta_0$ & Rescaled productivity before shocks, $z = K\zeta_0/\varepsilon$ (high-perishability limit) \\
$\zeta$ & Effective productivity $\zeta=\zeta_0 e^{-\sigma^2}$ \\
$\Xi_i(t)$ & Productivity increment, $\Xi_i(t)=\xi_i(t+1)-\xi_i(t)$ \\
$y_{<}(t)$ & Minimal production across suppliers of a given firm at time $t$ (large-economy limit) \\
$r_{<}(t)$ & Ratio $y_{<}(t)/\overline{y}_t$, measure of weakest-firm gap \\
$\gamma$ & Effective parameter $\gamma = \varkappa e^{-2\sigma\sqrt{\log K}}$ \\
$\sigma_c(\kappa)$ & Critical volatility (resilience–fragility boundary) at given $\kappa$ \\
$\kappa_c(\sigma)$ & Critical inventory level at given $\sigma$ \\
$\sigma_{\max}$ & Maximum volatility above which no inventory level ensures stability \\
$t_\times$ & Typical crash time (time to system-wide  collapse) \\
$\eta$ & Effective drift in the random evolution of $- \log r_<$ \\
$\Gamma$ & Effective volatility in the random evolution of $- \log r_<$ \\
\midrule
$R$ & Excess volatility ratio $R^2 = \mathbb{V}[\sum_i y_i]/\mathbb{V}[\sum_i z_i]$ \\
$\tau_c$ & Random time of first system-wide  crash \\
$\mathbb{P}_N[\tau_c \leq T]$ & Probability that an economy with $N$ firms crashes before time $T$ \\
$\chi(\sigma,T,N)$ & Variance of $\min(\tau_c,T)$ (crash-time variance) \\
$F,G,s,w$ & Scaling functions for crash probability and crash-time variance \\
\end{longtable}
\end{center}

\appendix
\numberwithin{equation}{section}
\renewcommand{\theequation}{\thesection\arabic{equation}}
\numberwithin{figure}{section}
\renewcommand{\thefigure}{\thesection\arabic{figure}}

\section{Cone-wise stability: linearised dynamics}
\label{app:cone-wise}

In this appendix, we assess the stability of the steady-state with respect to linear homogeneous perturbations. 
We write this perturbed solution as fluctuations around the stationary state, namely
\begin{equation*}
\left(\begin{matrix}
    y^\top(t) \\ y(t) \\ S(t) \\ O(t) \\ X(t) \\g(t)
\end{matrix}\right)= \left(
    \begin{matrix}
        y^\star+\Delta y^\top(t)\\ y^\star+\Delta y(t)\\S^\star+\Delta S(t)\\O^\star+\Delta O(t)\\ X^\star+\Delta X(t)\\ g^\star+\Delta g(t)
    \end{matrix} 
\right).
\end{equation*}

Because our system behaves differently depending on certain conditions -- for example, whether supply exceeds or falls short of demand -- our linear stability analysis must distinguish between these different different regions of state-space, or \textit{cones}. 
To this aim, we perform a \textit{cone-wise linear analysis}~\citep{Dessertaine2022, dessertaine2022non}.
Each cone has its own specific linearized dynamics, and a complete study of stability requires understanding not only the stability within each cones but also the switching dynamics between them.

First, we linearise dynamics \eqref{e1} to \eqref{e7} within each cone. This allows us to obtain the eigenvalues of the stability matrices within each cone and, hence, the intra-cone stability. 
Imaginary components in the eigenvalues indicate the presence of oscillations, while real components smaller (larger) than $1$ indicate local stability (instability), provided the dynamics do not leave the cone: as mentioned above, the complete stability analysis requires studying whether the dynamics stay or exit the cone under consideration.

We now set to derive the update rules for the perturbations $\Delta y$, $\Delta y^\top$, $\Delta S$, $\Delta O$, $\Delta X$ and $\Delta g$, from which we will then extract the stability matrix. 
When $O(t)<M(t)$, supply is sufficient to cater for orders, and, from Eq.~\eqref{e1} and Eqs.~\eqref{e3}--\eqref{e5}, we respectively obtain 
\begin{align}
\Delta y^\top(t)=&\Delta y(t)\;,&
\Delta X(t)=&\Delta O(t) = \frac{\kappa+1}{z}\Delta y(t)-\Delta S(t)\;.
\label{e13a}
\end{align}
Combining Eqs.~\eqref{e6} and \eqref{e13a}, we further get
\begin{eqnarray}
\Delta g(t+1)\!\!\!&=&\!\!\!\left[\Delta g(t)+\Delta y(t)-K\frac{\kappa+1}{z}\Delta y(t)+K\Delta S(t)\right](1-\psi). 
\label{e14a}
\end{eqnarray}
while Eqs.~\eqref{e7} and \eqref{e13a} yields 
\begin{eqnarray}
\Delta y(t+1)\!\!\!&=&\!\!\!\!(1-\omega)\,\Delta y(t)+\omega\left[K\frac{(\kappa+1)}{z}\Delta y(t)-K\Delta S(t)-\Delta g(t)\right].
\label{e15a}
\end{eqnarray}
Finally, Eqs.~\eqref{e2} and \eqref{e13a} gives
\begin{eqnarray}
\Delta S(t+1)\!\!\!&=&\!\!\!\frac{\kappa(1-\psi)}{z}\Delta y(t).
\label{e16a}
\end{eqnarray}
From Eqs.~\eqref{e13a} to \eqref{e16a}), we remark that only three quantities need to be updated in order to fully determine the system at the next time step: $\Delta y$, $\Delta g$ and $\Delta S$. 
However, it will prove to be more natural to characterize the update rules in terms of $\Delta a$, $ \Delta y$ and $\Delta S$, where $\Delta a$ is given by
\begin{equation}
  \Delta a := {{\left(1  - K\frac{\kappa+1}{z}\right) \Delta y+\Delta g+K\Delta S}}.  
\end{equation} 
The update rule Eqs.~(\ref{e14a}--\ref{e16a}) characterizing the evolution of the perturbations can then be recast in the following form
\begin{eqnarray}
\Delta a(t+1)&=&\left[K\frac{(\kappa+1)\,\omega}{z}+1-\omega-\psi\right]\Delta a(t)+\left(1-K\frac{1+\kappa\psi}{z}\right)\Delta y(t)\nonumber\;,\\
\Delta y(t+1) &=&\Delta y(t)-\omega\,\Delta a(t)\nonumber\;,\\
\Delta S(t+1) &=& \frac{\kappa(1-\psi)}{z}\Delta y(t).
\label{e20a}
\end{eqnarray}
The evolution of $\Delta a$ and $\Delta y$ is thus decoupled from $\Delta S$, allowing us to define a $2 \times 2$ stability matrix $\mathcal{L}^+$ as
\begin{eqnarray}
\left[\!\!\begin{array}{c}\Delta a(t+1)\\\\\\\Delta y(t+1)\end{array}\!\!\!\right]=\left[\begin{array}{cc}\displaystyle{1 + K\frac{(\kappa+1)\,\omega}{z}-\omega -\psi}&\quad\displaystyle{1-K\frac{1+\kappa\psi}{z}}\\\\-\omega&1\end{array}\right] \left[\!\!\!\begin{array}{c}\Delta a(t)\\\\\\\Delta y(t)\end{array}\!\!\right] :={\mathcal L}^+\left[\!\!\!\begin{array}{c}\Delta a(t)\\\\\\\Delta y(t)\end{array}\!\!\right]\;.
\label{e21a}
\end{eqnarray}
In Appendix~\ref{app:cone-conditions}, we show that the condition $O(t)<M(t)$ is equivalent to $\Delta a>0$. Hence, $\mathcal{L}^+$ describes the evolution of small perturbations within the positive cone $\Delta a > 0$: as soon as $\Delta a < 0$, the system switches cones and the evolution of perturbations is governed by yet another matrix $\mathcal{L}^-$ that we now set up to derive.
When $O(t)>M(t)$ (or equivalently $\Delta a<0$), supply is not sufficient to cater for orders within the supply chain and is more complicated. In this cone, the uniform perturbation takes the form
\begin{align}
\Delta y^\top(t) =& \Delta y(t) \label{e22a} \\
\Delta O(t) =& \frac{(\kappa+1)}{z}\Delta y(t)-\Delta S(t) \;,
\label{e23a}
\end{align}
which respectively stem from Eq.~\eqref{e1} and \eqref{e3}.
Linearizing Eq.~\eqref{e18a} and using \eqref{e23a} we then obtain
\begin{align}
\Delta X(t) := \Delta M(t) =& \Delta O(t)+\frac{1+\kappa\psi}{z}\Delta a(t) =\frac{\kappa+1}{z}\Delta y(t)-\Delta S(t)+\frac{1+\kappa\psi}{z}\Delta a(t)\;.
\label{e24a}
\end{align}
In this supply-limited cone, the household consumption $C(t)$ also has to be linearized in the form $C(t)=c+\Delta c(t)$ since it is rationed. We obtain that
\begin{align}
        \Delta c(t) =& \left[1-\frac{K(1+\kappa\psi)}{z}\right]\left[\Delta g(t)+\Delta y(t)-K\Delta O(t)\right]=\left[1-\frac{K(1+\kappa\psi)}{z}\right]\Delta a(t).
\label{e25a}
\end{align}
Using Eqs.~\eqref{e6}, \eqref{e18a} and \eqref{e25a} yields $\Delta g(t+1)=0$, as we expect for the negative cone.
Eqs.~\eqref{e7} and \eqref{e25a} gives the update rule for $\Delta y$ as
\begin{align}
        \Delta y(t+1) =&\Delta y(t) - \omega\Delta a(t) \;,
    \label{e27a}
\end{align}
while Eqs.~\eqref{e2} and \eqref{e27a} yields the update rule for $\Delta S $ 
\begin{eqnarray}
\Delta S(t+1) &\approx& \left[\frac{\kappa}{z}\Delta y(t)+\frac{1+\kappa\psi}{z}\Delta a(t)\right](1-\psi) \;.
\label{e28a}
\end{eqnarray}
Finally, combining Eqs.~\eqref{e27a}--\eqref{e28a} with $\Delta g(t+1)=0$, we get the evolution of $\Delta a $ as
\begin{equation}
\begin{split}
    \Delta a(t+1)\approx &\left[1-K\frac{1+\kappa\psi}{z}\right]\Delta y(t) -\left[\omega-K\,\frac{1+\omega+\kappa(\omega+\psi)-\psi(1+\kappa\psi)}{z}\right]\Delta a(t).
\end{split}
\label{e29a}
\end{equation}
In this cone as well, only two quantities are needed to determine the perturbation at the next time step: $\Delta a$ and $\Delta y$. Their evolution, as long as $\Delta a$ remains negative, reads
\begin{eqnarray}
\left[\!\!\begin{array}{c}\Delta a(t+1)\\\\\\\Delta y(t+1)\end{array}\!\!\!\right]=\mathcal{L}^-\left[\!\!\!\begin{array}{c}\Delta a(t)\\\\\\\Delta y(t)\end{array}\!\!\right].
\label{e30a}
\end{eqnarray}
with
\begin{eqnarray}
\mathcal{L}^-:=\left[\begin{array}{cc}\displaystyle \frac{K}{z}\big[1 + \omega(\kappa+1-z/K)-\psi(1+\kappa(\psi-1))\big]&\quad\displaystyle{1-K\frac{1+\kappa\psi}{z}}\\\\-\omega&1\end{array}\right].
\label{e30b}
\end{eqnarray}

\subsection{Cone conditions}\label{app:cone-conditions}
In this appendix, we present the two different cones which characterize the linearised dynamics of our supply chain.
These cones are defined by the respective conditions $M(t)\lessgtr O(t)$: in one cone production is limited by supplies while in the other it is limited by demand. 
From Eq.~\eqref{e3}, the perturbation of orders $\Delta O$ is obtained as 
\begin{align}
\label{eq:linearized_orders}
\Delta O(t) =& \frac{\kappa+1}{z}\Delta y(t)-\Delta S(t)\,
\end{align}
while the perturbation of the rationed quantity $\Delta M$ is given by Eq.~\eqref{e4} as
\begin{eqnarray}
\Delta M(t) &\approx& \frac{y^*\,c}{(KO^*+c)^2}\Delta O(t)+\frac{O^*}{KO^*+c}\left[\Delta y(t)+\Delta g(t)\right].
\label{e17a}
\end{eqnarray} 
Injecting the steady-state expression \eqref{eq:steady_state_eq1} and \eqref{eq:eqproduction} into \eqref{e17a} gives
\begin{eqnarray}
\Delta M(t)&\approx&\Delta O(t)+\frac{1+\kappa\psi}{z}\left[\Delta y(t)+\Delta g(t)-K\Delta O(t)\right]\;.
\label{e18a}
\end{eqnarray}
Eq.~\eqref{e18a} has a simple interpretation: $M(t) > O(t)$ ($M(t) < O(t)$) if the change in product inventories is greater (lesser) than the change in orders. 
Further using \eqref{eq:linearized_orders} into \eqref{e18a}, we obtain the equivalence between $M(t)\lessgtr O(t)$ and $\Delta a(t)\lessgtr 0$ as
\begin{align}
\Delta M(t) \approx &\Delta O(t)+\frac{1+\kappa\psi}{z}\underbrace{\left[\Delta y(t)+\Delta g(t)-K\frac{\kappa+1}{z}\Delta y(t)+K\Delta S(t)\right]}_{\Delta a(t)} .
\end{align}
Therefore, the two cases $M(t)\lessgtr O(t)$ put the system into two different ``cones'', depending on the sign of $\Delta a(t)$. We coined $\Delta a(t)>0$ ($\Delta a(t)<0$) as the positive (negative) cone.

\subsection{Linear stability in the $\omega \to 0$ limit}
\label{sec:small_omega}
In this appendix, we show that the steady-state \eqref{eq:eqproduction} always remains stable when the dynamics of the target production is slow, namely $\omega \to 0$, and perishability is weak, \textit{i.e.} $\psi \to 0$. 
More precisely, we assume that $\omega \to 0$ while $\psi = O(\omega)$ and study the linear stability of equilibrium against homogeneous perturbations in the two cones separately, starting with the demand-limited one.

\subsubsection*{The demand-limited cone $\Delta a > 0$}

Rescaling $\Delta y$ by $\sqrt{\omega}$, the linear operator $\mathcal{L}^+$ takes the form
\[
\mathcal{L}^+ = \mathbb{I} + \sqrt{\omega} \left[\begin{array}{cc}\displaystyle{\frac{K(\kappa+1)-z)}{z} \,\sqrt{\omega}-\frac{\psi}{\sqrt{\omega}}}&\quad\displaystyle{1-\frac{K}{z}}\\\\-1&0\end{array}\right] + O(\omega^{3/2}),
\]
leading to the following eigenvalues
\begin{equation}
    \lambda_\pm^+ = 1 \pm i \sqrt{\omega\left(1 - \frac{K}{z}\right)} + \frac12 \left[\omega \frac{ K(\kappa + 1) - z}{z} - \psi \right]+ O(\omega^{3/2}).  
\end{equation}
Note that $K/z < 1$ since we are studying the steady-state.
The leading term in $\sqrt{\omega}$ is imaginary, corresponding to oscillations, whereas the next order term is real and negative (i.e. stabilising) when \footnote{Actually, the following inequality does not require $\omega, \psi \to 0$ and holds generally.}
\begin{equation}\label{eq:positive_sector_unstable}
      z \left(1 +\frac{\psi}{\omega}\right) > K (\kappa + 1)
\end{equation}
Since equilibrium exists only provided $z > K(1+\kappa \psi)$ (see Eq. \eqref{eq:ss_cond_1}), we find that the demand-limited cone is always \textit{locally} stable when 
$\psi > \kappa \omega + O(\omega^{3/2})$, but may have an unstable direction otherwise. 

However, in spite of this intra-cone stability, the dynamics of the system will sooner or later push the variables into the supply-limited cone, which we now study.

\subsubsection*{The supply-limited cone $\Delta a < 0$}

The eigenvalues of $\mathcal{L}^-$ can also be computed to $O(\omega)$ in this case; one finds 
\begin{equation}
    \lambda_+^- = 1 - \omega + O(\omega^2)\qquad\mbox{and} \qquad \lambda_-^- = \frac{K}{z} + \frac{K(\kappa+1)\,\omega}{z}+ \frac{K(\kappa-1)\,\psi}{z}+O(\omega^2), 
\end{equation}
showing that both directions are stabilising in the supply-limited cone. 

\subsubsection*{Vector dynamics within each cone}
We now look at the angular dynamics of the perturbation vector $\vec{p}=(\Delta a, \Delta y)$ within the positive cone.
Defining the rescaled vector as $\vec{p}_r=(\Delta a, \sqrt{(1-K/z)/\omega}\Delta y)$, $\vec{p}_r$ is updated with a rescaled stability matrix $\mathcal{L}^+_r$ reading
\[
\mathcal{L}^+_r = \mathbb{I} + \sqrt{\omega\,(1-K/z)} \left[\begin{array}{cc}\displaystyle{0}&\quad\displaystyle{1}\\\\-1&0\end{array}\right] + O(\omega),
\]
which shows that in the limit $\omega\rightarrow 0$ and $\psi\sim O(\omega)$, $\mathcal{L}^+_r$ provides a pure rotation by an angle of magnitude $-\sqrt{\omega\,(1-K/z)}$. Hence starting from any positive initial angle $\theta_0$ in the plane $\Delta a > 0, \Delta y >0$, the rescaling $\Delta y \to \sqrt{(1-K/z)/\omega} \Delta y$ drives $\theta_0$ to $\pi/2 - O(\sqrt{\omega})$. 

Then after a number of time steps $\approx \pi/\sqrt{\omega\,(1-K/z)}$, that angle has rotated to $-\pi/2$ and the system leaves the positive cone and enters the negative cone $\Delta a < 0$. During that time, the eigenvalue $\lambda_+^+ > 1$ has made the norm of the initial vector grow by an amount $\propto (1 + O(\omega))^{\pi/\sqrt{\omega}} \sim e^{O(\sqrt{\omega})}$, which is still close to unity. 

We now discuss the angular dynamics of the perturbation vector $\vec{p}$ in the negative cone. To order $\omega$, we have
\[
\mathcal{L}^- = \left[\begin{array}{cc}\displaystyle{K/z}&\quad\displaystyle{1 - K/z}\\\\0&1\end{array}\right]  + O(\omega).
\]
We find that the evolution of the slope $\Theta:= \Delta y/\Delta a$ is given by
\begin{equation}
    \Theta_{t+1} = \frac{z\Theta_t}{K + (z - K) \Theta_t} + O(\omega),
\end{equation}
which has two stable fixed points: $\Theta^\star_1 =1 + O(\omega)$ and $\Theta_2^\star =  0 + O(\omega)$. 
This means that after entering the supply limited cone at $\theta = - \pi/2$, the angle quickly converges towards $-3\pi/4$ where the norm of the vector tends exponentially fast towards zero, with a speed limited by $\lambda_+^-$. In other words, equilibrium is restored after a  time $\sim 1/\omega$ which is indeed the natural time scale of the dynamics.

This concludes our analysis of linear cone-wise stability of equilibrium in the limit $\omega \to 0$: even when the demand-limited cone is unstable (i.e. when Eq. \eqref{eq:positive_sector_unstable} is not satisfied), the system necessarily switches to the supply-limited cone where it converges back towards equilibrium.

\subsection{Cone-wise linear stability: more general results}

In the case where $\omega$ is not small, a general analysis is more cumbersome. One can nevertheless establish that, for any $\omega, \psi$,
\begin{itemize}
    \item The demand-limited cone is unstable when Eq. \eqref{eq:positive_sector_unstable} is not satisfied, which can be rewritten as
    \begin{equation}
        \kappa > \kappa_c^+ := \frac{z\, (\omega + \psi)} {K\omega}-1
    \end{equation}
    i.e. when productivity $z$ is too low or when precautionary inventory $\kappa$ is too high.
    \item The supply-limited cone also becomes unstable when 
    \begin{equation}
        \kappa >  \kappa_c^- := \frac{z (1+ \omega) - K ( 1 + \omega - \psi)}{K(\omega + \psi - \psi^2)}.
    \end{equation}
    Note that when $\omega$ and $\psi$ tend to zero, this condition is never satisfied and we recover our previous result about the stability of the supply-limited cone.
\end{itemize}
Comparing these two conditions with the condition that an equilibrium exists, i.e., $\displaystyle{\kappa < \kappa_c^\star := ({z - K})/{(K \psi)}}$, one finds that the three transition points exactly coincide, i.e. $\kappa_c^\star = \kappa_c^- = \kappa_c^+$, when  
\[ \displaystyle{\frac{K}{z} = 1 - \frac{\psi^2}{\omega(1-\psi)}}.
\]
Thus, the two following regimes exist:
\begin{enumerate}
    \item When
      \[
    1 - \frac{\psi^2}{\omega(1-\psi)} < \frac{K}{z} < 1,
    \]
    one finds that $\kappa_c^\star < \kappa_c^- < \kappa_c^+$, i.e. dynamical instabilities appear in a region when equilibrium has already ceased to exist. Conversely, when $\kappa < \kappa_c^\star$, equilibrium is always stable. This is in particular the case when $\omega \leq \psi^2/(1-\psi)$, for example in the high perishability limit $\psi \to 1$, as confirmed in section \ref{sec:HP-MF} below.
    \item When 
    \[
    \frac{K}{z}  < 1 - \frac{\psi^2}{\omega(1-\psi)},
    \]
    $\kappa_c^+ < \kappa_c^- < \kappa_c^\star$. As $\kappa$ increases and reaches $\kappa_c^+$, instability occurs first in the demand-limited cone, while the supply-limited cone is still stable. Both cones are unstable (and hence equilibrium must be unstable) when $\kappa > \kappa_c^-$. {Note that when $\psi=O(\omega)$ and $\omega \to 0$, one has that $\kappa_c^- \to \infty$, recovering the results of previous section \ref{sec:small_omega}.}
\end{enumerate}
We have unfortunately not been able to find simple general conditions for global stability when $\kappa$ is in the interval $[\kappa_c^+,\kappa_c^-]$, i.e. when the supply-limited cone is stable and the demand-limited cone unstable. Such conditions require the analysis of the angular dynamics in the two cones. While the limit $\omega \to 0$ allowed us to conclude that equilibrium is always stable, it appears that for larger $\omega$ the angular dynamics may not get trapped anymore in the (stable) $\Delta a < 0$ cone and the discussion of global stability becomes much more subtle, as it could result from a balance between the time spent in the unstable (demand-limited) region and in the stable (supply-limited) region -- see \citet{dessertaine2022non} for a similar situation. Numerical simulations reveal an even more complex structure, where the system actually undergoes a Hopf-type bifurcation (a threshold at which the steady state gives way to persistent oscillations) for $\kappa$ inside the interval $[\kappa_c^+,\kappa_c^-]$.
The latter destabilizes the fixed point towards a non-linear steady periodic orbit. The regime diagram in the $(\psi, \kappa)$ plane is displayed on Fig. \ref{fig:unstability_phase_diagram}  for $K/z=1/3$ and three different values of $\omega$. It shows that the oscillating region occupies a larger and larger fraction of the $[\kappa_c^+,\kappa_c^-]$ interval as $\omega$ increases.

In the following, we will avoid these complications and stay firmly within the linearly stable region, defined by 
\begin{equation}
    \kappa_{\min} := \frac{1}{1-\psi} < \kappa < \min(\kappa_c^\star, \kappa_c^+),
\end{equation}
and study the possibly destabilizing role of productivity shocks, mediated by failure cascades.

\section{Finite-size scaling analysis of the resilience-to-fragility transition}\label{app:finite-size}
In this appendix, we characterize the behavior of the resilient-to-fragile transition observed in numerical simulation with the tools of statistical physics.
It is consistent with a genuine continuous phase transition as there is no discrete jump in crash probability; the economy becomes progressively more fragile as $\sigma$ approaches a critical value. 
We characterize it using a convergence analysis for finite economies (finite-size scaling in statistical physics), following the approach of \citet{fosset2020endogenous} for a model of {\it liquidity crises} in financial markets. 
The scenario we want to test is that, for a given $\kappa$ and in the limit $N\to \infty$, there exists a critical value of the noise amplitude $\sigma$ such that for $\sigma < \sigma_c(\kappa)$, an infinite-size economy never crashes, while for $\sigma > \sigma_c(\kappa)$ it eventually crashes with probability one. 

In order to investigate such a scenario, we study the probability $\mathbb{P}_N(\tau_c < T)$ that an economy with $N$ firms crashes at a time $\tau_c$ less than a given $T$. 
In fact, Fig  \ref{fig:phase_diagram}-a shows a heat map of such a probability for different values of $(\sigma, \kappa)$. 
Although suggestive, this plot cannot be used to conclude on the existence of a genuine transition separating a crisis-safe from a crash-prone regime, provided one waits long enough. 
Mathematically, the question is about the subtle behavior of $\mathbb{P}_N(\tau_c < T)$ in the double limit $N,T \to \infty$. 

\begin{figure}
    \hspace*{-0.3cm}
    \begin{tikzpicture}
    \node at (0,0) {\includegraphics{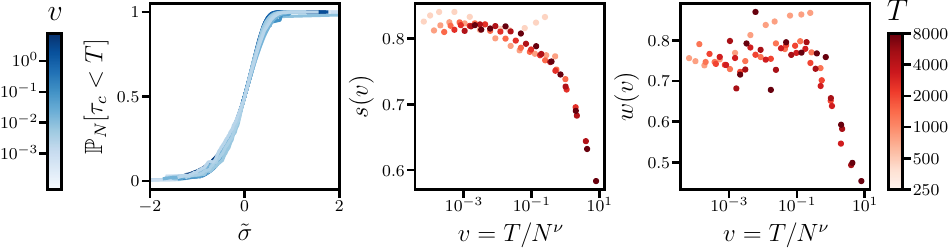}};
    \tikzmath{\dy=-0.8 ; \ss=1.1;}
    \node[font=\bf,scale=\ss] at (-2.7,\dy) {a.};
    \node[font=\bf,scale=\ss] at (-0.6,\dy) {b.};
    \node[font=\bf,scale=\ss] at (3.9,\dy) {c.};
    \end{tikzpicture}
    \caption{{\bf a.} Cumulative distribution $\mathbb P_N[\tau_c \leq T]$ as a function of the rescaled variable $\tilde{\sigma}=\ell^2(\sigma-\sigma_m(v))/w(v)$ for various values of $T$ and $N$. $\sigma_m(v)$ is determined numerically by inverting $\mathbb P_N[\tau_c \leq T](\sigma_m)=1/2$ while $\ell^{-2}w(v)$ is determined from the slope of $\mathbb P_N[\tau_c \leq T](\sigma)$ at $\sigma_m$. {\bf b.} Numerically determined scaling function $s(v)=\ell^2(\sigma_m(v)-\sigma_c)$ as a function of $v$. {\bf c.} Numerically determined scaling function $w(v)=\ell^2 W(v)$, with $W(v)$ the numerical width of $\mathbb P_N[\tau_c \leq T]$ at fixed $T$ and $N$. Note that outliers correspond to small values of $N,T$, for which deviations are expected. Parameters: $c=6$, $\omega=\psi=0.1$, $K=6$, $z=18$, $L_0=1$, $\kappa=2.6$.}
    \label{fig:cumul}
\end{figure}

On one hand, if the limit $T \to \infty$ is taken first, one expects that there is always a non zero probability (although very small) that the economy crashes, due to rare, very large shocks which surely happen in the long run. Hence for a given $\kappa$, one has
\[
\lim_{N \to \infty} \lim_{T \to \infty} \mathbb P_N[\tau_c \leq T,\sigma] = 1\, , \quad \forall \sigma\, .
\]
On the other hand, if the limit $N \to \infty$ is taken first, one may be in a situation where, for a fixed value of $\kappa$,
\begin{equation}
\lim_{T \to \infty} \lim_{N \to \infty} \mathbb P_N[\tau_c \leq T,\sigma] = \begin{cases} 1 \quad \text{when} \quad  \sigma > \sigma_c(\kappa), \\
0 \quad \text{when} \quad  \sigma < \sigma_c(\kappa).
\end{cases}
\end{equation}
Since numerical simulations can only be done for finite $N$ and $T$, a common strategy is to use finite size scaling to extrapolate to infinite sizes and waiting times \citep{binder1997applications}. Focusing on the slow, low perishability case $\omega=\psi=0.1$, with $K=6$, $z=18$ and $\kappa=2.6$, we found the following behavior to hold for large enough $N$ and $T$,
\begin{equation}\label{finitesize}
\mathbb P_N[\tau_c \leq T,\sigma] \approx F\Big( \frac{\sigma - \sigma_m(v)}{\ell^{-2} \,w(v)}\Big), \qquad  \sigma_m(v) = \sigma_c + \ell^{-2} s(v),  
\end{equation}
as shown on Fig. \ref{fig:cumul}. In Eq. \eqref{finitesize}, $\ell:=\log (aN)$, $a \approx 3.75$, $v := T/N^{\nu}$, $\nu \approx 3/2$, $\sigma_c \approx 0.7833$. 
The exponent $\nu$ governs how the transition sharpens with system size: it determines the relative rate at which the simulation horizon $T$ must grow with the number of firms $N$ for the transition to be observable. 
The function $F(u)$ is a monotonic regular function increasing from $F=0$ when $u \to -\infty$ to $F=1/2$ for $u=0$ and to $F=1$ for $u \to + \infty$. The functions $w(v)$ and $s(v)$, shown in Fig. \ref{fig:cumul}, respectively govern the scaling behavior of the width and location of the transition. 
Note that $w_0:=w(0) \approx 0.18$ and $s_0:=s(0) \approx 0.825$ (corresponding to the regime $N^{\nu} \gg T$) are both finite and non-zero. A similar scaling form was found to hold in the case of a limit order book model for liquidity crises by \citet{fosset2020endogenous}. 

\begin{figure}
    \begin{tikzpicture}
    \node at (0,0) {\includegraphics[width=0.5\linewidth]{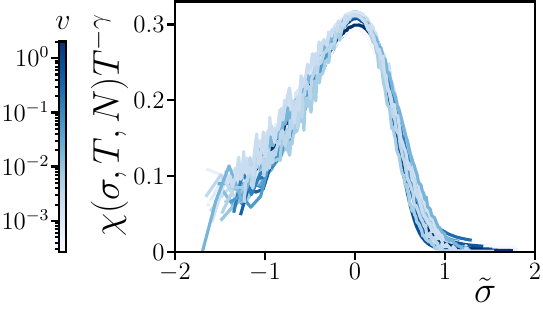}};
    \node at (8,0) {\includegraphics[width=0.5\linewidth]{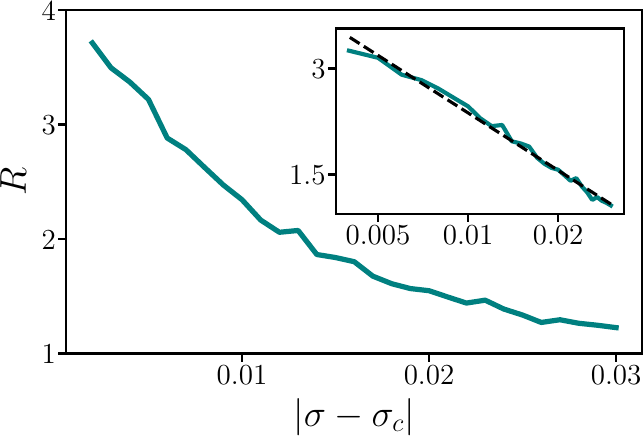}};
    \tikzmath{\dy=2;\ss=1.3;}
    \node[scale=\ss] at (-1,\dy) {a.};
    \node[scale=\ss] at (6.1,\dy) {b.};
    \end{tikzpicture}
    \caption{{\bf a.} Scaling plot of the rescaled crash-time variance $\chi(\sigma,T,N)T^{-\gamma}$ as a function of $\tilde{\sigma}=(\sigma-\sigma_m(v))/\ell^{-2}w(v)$ with exponent $\gamma=1.05$. {\bf b.} Excess volatility $R$ (defined in Eq. \eqref{eq:excess_vol}) as a function of $\sigma-\sigma_c$ for the cross-section at $\kappa=2.6$ (teal line in \Fref{fig:phase_diagram}-a). (Inset) Log-log plot of $R$ vs. $\sigma - \sigma_c$, consistent with an inverse-square-root divergence $R \sim (\sigma_c - \sigma)^{-1/2}$. The dashed line indicates a slope of $-0.54$. Parameters: $c=6$, $\psi=0.1$, $K=6$, $z=18$, $L_0=6$, $T=2000$, $N=750$, $\omega=0.1$, $\sigma_c=0.8$.}
    \label{fig:excess_vol}
\end{figure}

We furthermore expect -- as is standard for continuous transitions in statistical physics \citep{binder1997applications,sethna2021statistical} -- that the scaling behavior in Eq. \eqref{finitesize} is robust and qualitatively independent of $\kappa, K, z, \psi, \omega$ as well as many other details of the model (although the numerical values of $\sigma_c, a, s_0$ and $w_0$ will depend on model parameters). This robustness -- called {\it universality} in statistical physics -- implies that the qualitative nature of the transition depends only on broad features of the model rather than on specific parameter values. 
Note in particular that
\begin{enumerate}
    \item Since $0 < w_0 < +\infty$ (see Fig. \ref{fig:cumul}-c), the width over which $\mathbb P_N[\tau_c \leq T,\sigma]$ transitions shrinks as $\ell^{-2} = 1/\log^{2}N$ in the limit $N \to \infty$. 
    \item For $T \gg N^{\nu}$, on the other hand, the width becomes independent of $N$ but decreases with $T$.  
    \item Since $0 < s_0 < +\infty$ (see Fig. \ref{fig:cumul}-b), $\sigma_c$ can be identified as the critical shock amplitude beyond which very large economies surely crash after a finite time.  
    \item $s(v)$ is decreasing with $v$ and we conjecture that $s(v)\to -\infty$ when $v \to \infty$. Such a behavior is compatible with a failing economy when $T$ becomes extremely large for fixed values of $N$ and $\sigma$.
    \item When $N^{\nu} \gg T$, both the deviation $\sigma_m - \sigma_c = s_0 \ell^{-2} $ and the width $w_0 \ell^{-2} $ are of the same order of magnitude. This is often the case for finite-economy corrections close to continuous transitions \citep{binder1997applications}.  
\end{enumerate}
Another common method of determining the values of $\sigma_c$ is to study the variance of the first crisis time \citep{binder1997applications}, defined as
\[
\chi(\sigma,T,N) = \mathbb{V} \left[ \min(\tau_c, T) \right] 
\]
for a fixed value of $\kappa$ and different values of $\sigma, T$ and $N$. This quantity -- which measures the sensitivity of the crash time to parameter changes, is analogous to the {\it susceptibility} in statistical physics. We expect it to peak close to the transition boundary, since for small $\sigma$, $\tau_c$ is nearly always larger than $T$ implying $\chi \to 0$, whereas for large $\sigma$, the crash time $\tau_c$ is always small and therefore $\chi$ is also small. The finite size scaling assumption for this quantity amounts to:
\begin{equation}
   \chi (\sigma, T,N) = T^{\gamma}\, G\Big( \frac{\sigma - \sigma_m(v)}{\ell^{-2} \,w(v)}\Big),
\end{equation}
where $\sigma_m$ is given by Eq.~\eqref{finitesize}, $\gamma$ is a new exponent and $G(u)$ is a humped function that goes to zero for $u \to \pm \infty$ \citep{fosset2020endogenous}. We show in Fig. \ref{fig:excess_vol}-a the corresponding scaling plot for different values of $T$ and $N$, using the very same scaling functions $s(v)$ and $w(v)$ as in Fig. \ref{fig:cumul}-b,c. The shape of $G(u)$ conforms to expectations: it peaks close to $u=0$ and tends to zero at $\pm \infty$. The fact that data for different $(T,N)$ fall onto a single curve further supports the continuous nature of the resilient-to-fragile transition in our model.

Another quantity of immediate interest is the excess volatility ratio $R$, defined as  
\begin{equation}\label{eq:excess_vol}
    R^2:= \frac{\mathbb{V}[\sum_i y_i ]}{\mathbb{E}[\sum_i y_i ]^2} \, \frac{\mathbb{E}[\sum_i z_i ]^2}{\mathbb{V}[\sum_i z_i ]}\;.
\end{equation}
$R$ measures the contribution of network effects in the volatility of aggregate production: $R < 1$ when amplification effects can be neglected, i.e. when $\sigma \to 0$ and $\kappa \gg 1$. This ratio is reported in Fig. \ref{fig:excess_vol}-b, showing that excess volatility increases as $\sigma$ approaches $\sigma_c$, and in fact diverges as $\sim (\sigma_c - \sigma)^{-1/2}$, see Fig. \ref{fig:excess_vol}-b.

Finally, note that $\kappa_c(\sigma)$ appears to diverge for a certain value $\sigma_{\max}$, beyond which the economy is always unstable, regardless of the level of precautionary inventories. We will see below that our large-economy approximation indeed predicts such a threshold, with a value of 
$\sigma_{\max}$ given by Eq. \eqref{eq:sigma_max} in the high perishability limit.

\section{The high perishability limit}\label{app:high-perishability}
In this appendix, we establish the update rules \eqref{eq:limit_prod} to \eqref{eq:limit_plan} in the high-perishability limit where $\psi = 1 - \varepsilon$ and $\varepsilon \to 0$.
To this aim, we start from the evolution of the model Eqs. \eqref{e1} to \eqref{e7} and linearise them at first non-vanishing order in $\varepsilon$.
The inventory dynamics of input good $j$ at firm $i$ reading
\begin{equation}
S_{ij}(t+1)
=
\left[S_{ij}(t)-\frac{y_i(t)}{z_i(t)}+X_{ij}(t)\right](1-\psi)\;,
\end{equation}
becomes
\begin{equation}
S_{ij}(t+1)
=
\varepsilon\left[S_{ij}(t)-\frac{y_i(t)}{z_i(t)}+X_{ij}(t)\right].
\end{equation}
In the scaling introduced in Section~\ref{sec:HP-MF}, both $S_{ij}$ and the
ratio $y_i/z_i$ are of order $\varepsilon$, whereas the flows $X_{ij}$ are of
order one. Writing
\[
S_{ij}(t) = \varepsilon\, S_{ij}^{(0)}(t) + O(\varepsilon^2), \qquad
X_{ij}(t) = X_{ij}^{(0)}(t) + O(\varepsilon), \qquad
\frac{y_i(t)}{z_i(t)} = O(\varepsilon),
\]
we obtain, to first order in $\varepsilon$,
\begin{equation}
S_{ij}(t+1)
=
\varepsilon\, X_{ij}^{(0)}(t) + O(\varepsilon^2).
\end{equation}
Thus, in the high-perishability limit, input inventories are determined by
the contemporaneous flows and simply reflect last period's deliveries at order
$\varepsilon$.

We then consider the orders of downstream firms to their suppliers, namely
\begin{equation}
O_{ij}(t)=\left[(\kappa+1)\frac{y^\top_i(t)}{z_i(t)}-S_{ij}(t)\right]^+.
\end{equation}
Using the scalings $\kappa = \varkappa/\varepsilon$ and
$z_i(t) \sim K\zeta_0 e^{\xi_i(t)}/\varepsilon$ within this high-perishability limit, the combination
$(\kappa+1)y^\top_i/z_i$ remains of order one, while $S_{ij}(t)$ is of order
$\varepsilon$. Therefore, at leading order, $O_{ij}$ is given by
\begin{equation}
O_{ij}(t) \approx \varkappa\,\frac{y^{\top(0)}_i(t)}{z_i(t)}
\equiv O_{ij}^{(0)}(t).
\end{equation}
Hence, to first order in $\varepsilon$, orders are pinned down by target production
and productivity.
Similarly, the inventory of finished goods evolves according to
\begin{equation}
g_i(t+1)
=
\left[g_i(t)+y_i(t)-\sum_{j \in {\mathcal{C}}(i)} X_{ji}(t)-C_i(t)\right](1-\psi),
\end{equation}
which, in the high-perishability limit, becomes
\begin{equation}
g_i(t+1)
=
\varepsilon\left[g_i(t)+y_i(t)-\sum_{j \in {\mathcal{C}}(i)} X_{ji}(t)-C_i(t)\right].
\end{equation}
Again, $g_i(t)$ is of order $\varepsilon$, so to leading order we obtain
\begin{equation}
g_i(t+1)
=
\varepsilon\left[y_i^{(0)}(t)-\sum_{j \in {\mathcal{C}}(i)} X_{ji}^{(0)}(t)-C_i(t)\right]
+ O(\varepsilon^2).
\end{equation}
Thus, both input inventories $S_{ij}$ and finished-goods inventories $g_i$ are
of order $\varepsilon$ and are determined by the order and production flows in this
limit, as one would expect when goods depreciate almost instantaneously.
The target production rule is given by
\begin{equation}
y^\top_i(t+1)= (1 - \omega) y^\top_i(t) + \omega \left[\min\left(
\left[c_i + \sum_{j \in {\mathcal{C}}(i)}O_{ji}(t)-g_i(t)\right]^+,
z_i(t)\min_{j\in \mathcal{S}(i)}S_{ij}(t),
z_i(t)L_0
\right)\right].
\end{equation}
The $\min$ is
dominated by its first two arguments $z_i(t)S_{ij}(t) = O(1)$ and $z_i(t)L_0= O(\varepsilon^{-1})$ so that, to leading order, target production is updated as
\begin{align}
y^{\top (0)}_i(t+1) &= (1-\omega) y^{\top (0)}_i(t) + \omega \min\left(c+\frac{\varkappa}{\zeta}\overline{y}^\top_i(t),\min_{j \in \mathcal{S}(i)}\overline{x}_{ij}(t)\right)\;,
\end{align}
where $\overline{y}^\top_i(t) := K^{-1} \sum_{j\in \mathcal{C}(i)} {y^{\top (0)}_j(t)}$ and $\overline{x}_{ij}(t) := K\zeta_0 e^{\xi_i(t)} X_{ij}^{(0)}(t)$.
Finally, the production function
\begin{equation}
y_i(t)
=
\min\left[y^\top_i(t),z_i(t)\, \min_{j\in{\cal S}(i)} S_{ij}(t),z_i(t)L_0\right],
\end{equation}
reduces at leading order, using $S_{ij}(t) \approx \varepsilon X_{ij}^{(0)}(t)$
and the rescaling $z_i(t) = K\zeta_0 e^{\xi_i(t)}/\varepsilon$, to
\begin{equation}
y_i^{(0)}(t)
=
\min\left[
y_i^{\top(0)}(t),
\, e^{\xi_i(t)-\xi_i(t-1)} \, \min_{j\in{\cal S}(i)} \overline{x}_{ij}(t-1)
\right],
\end{equation}
which is of order one. This is
precisely Eq.~\eqref{eq:limit_prod} in the main text.
In summary, at first non-vanishing order in $\varepsilon$ (high perishability),
the inventory variables $S_{ij}$ and $g_i$, as well as the orders $O_{ij}$, are
algebraically determined by production and target production. The effective
dynamics is fully captured by the rescaled variables
$\overline{x}_{ij}(t)$, $y_i(t)$, and $y_i^\top(t)$, as given by
Eqs.~\eqref{eq:limit_prod}–\eqref{eq:limit_plan}.

\section{Derivation of the critical line in the high-perishability limit}
\label{app:HP-derivation}

This appendix provides the full derivation of the critical shock 
amplitude $\sigma_c$ stated in Eq.~\eqref{eq:sigmac_main} as well as the 
crash-time formula \eqref{eq:crash_time_main}. The starting point is 
the rescaled dynamics \eqref{eq:limit_prod}--\eqref{eq:limit_plan} 
obtained in Section~\ref{sec:HP-numerical}; the derivation of those 
equations from the original model is given in 
Appendix~\ref{app:high-perishability}.

\subsection{Steady state}
\label{app:HP-steady}

We begin by identifying the homogeneous steady state of 
\eqref{eq:limit_prod}--\eqref{eq:limit_plan} in the absence of shocks. 
We make two assumptions, both of which we verify self-consistently 
below: (i) realised production matches its target, $y_i = y^\top_i$, so 
the bottleneck term in \eqref{eq:limit_prod} is not binding; and (ii) 
the household-consumption term in the target update rule 
\eqref{eq:limit_plan} dominates the bottleneck term, so the target production
adjusts toward $c + (\varkappa/\zeta)\overline{y}^\top_t$.
Under these assumptions, \eqref{eq:limit_plan} becomes
\begin{equation}
\label{eq:app_prod_update}
    y^\top_i(t+1) = (1-\omega)\, y^\top_i(t) + \omega\left(c + \frac{\varkappa}{\zeta}\, \overline{y}^\top_i(t)\right).
\end{equation}
In the homogeneous case $y^\top_i = \overline{y}^\top_i(t) = y^\star$, the 
fixed point of this equation is
\begin{equation}\label{eq:app_ystar}
    y^\star = c + \frac{\varkappa}{\zeta}\, y^\star \;\Longrightarrow\; y^\star = \frac{\zeta\, c}{\zeta - \varkappa},
\end{equation}
which exists and is positive whenever $\zeta > \varkappa$. This is the 
analogue, in the rescaled variables, of the Hawkins--Simon condition 
\eqref{eq:ss_cond_1}. Recalling that $\zeta = \zeta_0\, e^{-\sigma^2}$, 
the condition can equivalently be written
\begin{equation}\label{eq:app_HS_noisy}
   \zeta_0 > \varkappa\, e^{\sigma^2}.
\end{equation}
For fixed average productivity $\bar\zeta$, this places an upper bound 
on the shock amplitude $\sigma$ beyond which the steady state ceases to 
exist; this is the origin of $\sigma_{\max}$ derived in 
\ref{app:HP-sigma-max} below.

We verify assumption (ii) at the steady state. The two arguments of the 
$\min$ in \eqref{eq:limit_plan} are $c + (\varkappa/\zeta) y^\star = y^\star$ 
and $\min_j \overline{x}^\star_{ij} = \varkappa\, y^\star$ (using 
\eqref{eq:limit_x} at the steady state, where the $\min$ inside 
\eqref{eq:limit_x} equals one). Since $\varkappa > 1$, the bottleneck 
term exceeds the consumption term, and the consumption term is selected. 
Assumption (i) follows from \eqref{eq:limit_prod}: at the steady state, 
$\overline{x}^\star_{ij} = \varkappa\, y^\star > y^\star$, so the 
bottleneck does not bind.

\paragraph{Local stability.} Linearising \eqref{eq:app_prod_update} around the homogeneous solution $y_i^\top (t) = \overline{y}_i^\top (t) = y^\star +\Delta y(t) $ gives
\begin{equation}
    \Delta y(t+1) = \left[1 - \omega + \omega\, \frac{\varkappa}{\zeta}\right] \Delta y(t),
\end{equation}
with contraction rate $1 - \omega(1 - \varkappa/\zeta) < 1$ when 
$\varkappa < \zeta$. The steady state is therefore locally stable to 
homogeneous perturbations whenever it exists, confirming the claim 
made in Section~\ref{sec:HP-critical}.

\subsection{Two coupled dynamics}
\label{app:HP-coupled}

We now reintroduce shocks and characterize the joint dynamics of the 
average target production and the worst supplier's output. Because the 
random regular network is locally tree-like, in the large-$K$ limit the 
production levels at neighbouring firms are approximately independent, 
and we can describe their distribution by a single density $P(y, t)$. 
Two functionals of this distribution will play distinct roles:
\begin{itemize}
    \item the average $\overline{y}^\top_t := \int y^\top P(y^\top, t)\, dy^\top$, capturing the bulk of the distribution;
    \item the typical minimum over $K$ samples, $y_<(t)$, capturing the left tail.
\end{itemize}

Under our assumption of fast-decaying $P(y, t)$, the distribution of 
the minimum of $K$ i.i.d.\ samples drawn from $P$,
\begin{equation}
    Q(y_m, t) = K\, P(y_m, t) \left[\int_{y_m}^\infty P(y, t)\, dy\right]^{K-1},
\end{equation}
peaks sharply for large $K$ around an $i$-independent value $y_<(t)$. 
This concentration property is what allows the homogeneous-network 
description below to remain self-consistent.

\subsection{Dynamics of the average $\overline{y}^\top_t$}
\label{app:HP-avg}

Let us first recall the starting point equations:
\begin{align}
    y_i(t) &= \min\left[y^\top_i(t),e^{\xi_i(t)-\xi_i(t-1)} \, \min_{j \in \mathcal{S}(i)} \overline{x}_{ij}(t-1)\right], \label{eq:limit_prod2} \\
    \overline{x}_{ij}(t) &= \varkappa y^\top_i(t) \min\left[1 ,\displaystyle{\frac{\zeta \,  y_j(t)}{\varkappa \overline{y}^\top_i(t)+{\zeta}c}}\right], \label{eq:limit_x2} \\
    y^\top_i(t+1) &= (1-\omega) y^\top_i(t) + \omega \min\left(c+\frac{\varkappa}{\zeta}\overline{y}^\top_i(t),\min_{j \in \mathcal{S}(i)}\overline{x}_{ij}(t)\right). \label{eq:limit_plan2}
\end{align}

Now, provided the inner $\min$ in \eqref{eq:limit_x2} is attained at the 
non-trivial argument, i.e.\ $\zeta\, y_<(t) < \varkappa\, \overline{y}^\top_t + \zeta\, c$, we can link the minimal flow to firm $i$ to the 
weakest production $y_<(t)$ as
\begin{equation}\label{eq:app_min_x}
    \min_{j \in \mathcal{S}(i)} \overline{x}_{ij}(t) = y^\top_i(t)\, \frac{\varkappa\, \zeta\, y_<(t)}{\varkappa\, \overline{y}^\top_t + \zeta\, c}.
\end{equation}
Inserting \eqref{eq:app_min_x} into the target update rule 
\eqref{eq:limit_plan2}, we obtain
\begin{equation}\label{eq:app_yhat_full}
    y^\top_i(t+1) = (1-\omega)\, y^\top_i(t) + \omega\, \left(c + \frac{\varkappa}{\zeta}\, \overline{y}^\top_t\right) \min(1, \Pi_t),
\end{equation}
where
\begin{equation}
    \Pi_t := \varkappa\, \frac{\zeta\, y^\top_i(t)}{\varkappa\, \overline{y}^\top_t + \zeta\, c}\, \frac{\zeta\, y_<(t)}{\varkappa\, \overline{y}^\top_t + \zeta\, c}.
\end{equation}
For $\omega \ll 1$, target productions remain close to their average, 
so $y^\top_i(t) \approx \overline{y}^\top_t$ and $\Pi_t$ becomes 
$i$-independent.
It is convenient to introduce the function
\begin{equation}\label{eq:app_f_def}
    f(y) := \frac{\zeta\, y}{\varkappa\, y + \zeta\, c},
\end{equation}
which is increasing in $y$ and satisfies $f(y^\star) = 1$. 
Using this definition, $\Pi_t$ reads
\begin{equation}\label{eq:app_Pi_simplified}
    \Pi_t = \varkappa\, r_<(t)\, f\!\left(\overline{y}^\top_t\right)^2\;,
\end{equation}
where $r_<(t) := y_<(t)/\overline{y}^\top_t$.
Equation \eqref{eq:app_yhat_full} has two 
regimes depending on whether $\Pi_t$ exceeds unity.

\textit{Regime 1: $\Pi_t > 1$.} The $\min$ in \eqref{eq:app_yhat_full} 
selects the consumption term, and the dynamics reduces to
\begin{equation}\label{eq:app_avg_regime1}
    \overline{y}^\top_{t+1} = (1-\omega)\, \overline{y}^\top_t + \omega\left(c + \frac{\varkappa}{\zeta}\, \overline{y}^\top_t\right),
\end{equation}
which converges back toward $y^\star$ whenever $\varkappa < \zeta$ --- 
i.e., whenever the steady state exists. The bottleneck does not bind.

\textit{Regime 2: $\Pi_t < 1$.} The $\min$ selects the bottleneck term, 
and \eqref{eq:app_yhat_full} becomes
\begin{equation}\label{eq:app_avg_regime2}
    \overline{y}^\top_{t+1} = \overline{y}^\top_t \left[1 - \omega + \omega\, \varkappa\, r_<(t)\, f\!\left(\overline{y}^\top_t\right)\right],
\end{equation}
which is the equation \eqref{eq:avg_dynamics} of the main text. In this 
regime, the average decays unless the bracketed factor exceeds one, 
which requires $\varkappa\, r_<(t)\, f(\overline{y}^\top_t) > 1$.

\paragraph{When does the bottleneck bind?} A necessary condition for 
the $\overline{y}_t^\top$ to decay is $\Pi_t < 1$, which using 
\eqref{eq:app_Pi_simplified} reads $\varkappa\, r_<\, f^2 < 1$. Since 
$f(\overline{y}^\top_t) \leq 1$ for $\overline{y}^\top_t \leq y^\star$ 
and $r_< \leq 1$ by definition, this can be satisfied even when 
$\varkappa > 1$, provided $r_<$ is small enough. The fate of the 
average is therefore tied to the dynamics of the worst supplier, which 
we examine next.

\subsection{Dynamics of the worst supplier $y_<$}
\label{app:HP-worst}

To close the system, we need an evolution equation for $y_<(t)$. We 
return to the production equation \eqref{eq:limit_prod2}, which we 
rewrite using
\begin{equation}
    \Xi_i(t) := \xi_i(t+1) - \xi_i(t),
\end{equation}
a Gaussian variable of zero mean and variance $2\sigma^2$ since the 
$\xi_i$ are independent gaussian random variable themselves. Substituting \eqref{eq:app_min_x} into 
\eqref{eq:limit_prod2} and using the homogeneity of target productions, we obtain
\begin{equation}
    y_i(t+1) = y^\top_i(t+1)\, \min\!\left[1,\; \varkappa\, e^{\Xi_i(t)} \min\!\left(1,\; r_<(t)\, f\!\left(\overline{y}^\top_t\right)\right)\right].
\end{equation}
For $\overline{y}^\top_t \leq y^\star$ we have $f(\overline{y}^\top_t) \leq 1$ 
and $r_<(t) \leq 1$, so the inner $\min$ selects $r_<(t)\, f(\overline{y}^\top_t)$, 
and
\begin{equation}\label{eq:app_yi_realized}
    y_i(t+1) = y^\top_i(t+1)\, \min\!\left[1,\; \varkappa\, e^{\Xi_i(t)}\, r_<(t)\, f\!\left(\overline{y}^\top_t\right)\right].
\end{equation}
Taking 
the minimum over the $K$ suppliers in \eqref{eq:app_yi_realized}:
\begin{equation}\label{eq:app_ymin_full}
    y_<(t+1) \approx \overline{y}^\top_{t+1}\, \min\!\left(1,\; \varkappa\, \min_{i=1,\ldots,K} e^{\Xi_i(t)}\, r_<(t)\, f\!\left(\overline{y}^\top_t\right)\right).
\end{equation}

\paragraph{Extreme-value asymptotics.} For $K$ i.i.d.\ Gaussian variables $\{\Xi_i\}_{i\in[1,K]}$ with variance $2\sigma^2$, the 
typical value of the minimum scales as
\begin{equation}\label{eq:app_min_gaussian}
    \min_{i=1,\ldots,K} \Xi_i \approx -2\sigma\sqrt{\log K}
\end{equation}
when $K$ is sufficiently large. Note that a more 
refined statement, including the $O(1)$ Gumbel correction, is needed 
for the volatility calculation discussed in section \ref{sec:excess_vol}.
Substituting \eqref{eq:app_min_gaussian} into \eqref{eq:app_ymin_full} 
and dividing both side by $\overline{y}^\top_{t+1}$ gives
\begin{equation}\label{eq:app_r_evolution}
    r_<(t+1) = \min\!\left(1,\; \gamma\, r_<(t)\, f\!\left(\overline{y}^\top_t\right)\right), \qquad \gamma := \varkappa\, e^{-2\sigma\sqrt{\log K}}.
\end{equation}
For $\gamma < 1$ and $f(\overline{y}^\top_t) \leq 1$, the $\min$ is  saturated by its non-trivial argument, and we recover 
\eqref{eq:r_dynamics}
\begin{equation}
\label{eq:r_f_dynamics}
    r_<(t+1) = \gamma\, r_<(t)\, f\!\left(\overline{y}^\top_t\right).
\end{equation}

\subsection{The transition}
\label{app:HP-transition}

Close to the steady-state $f(\bar{y}_t^\top)\simeq 1$ and will start to decay only when $\varkappa r_{<} < 1$ as shown in Appendix \ref{app:HP-avg}.
Therefore, in the initial regime preceding the collapse equation \eqref{eq:r_f_dynamics} further simplifies into the multiplicative update
\begin{equation}
\label{eq:r_simple_dynamics}
    r_<(t+1) = \gamma\, r_<(t)\;.
\end{equation}
Linearising around the steady state $(y^\star, 1)$, where 
$f(y^\star) = 1$, the system decouples to leading order:
\begin{align}
    \overline{y}^\top_{t+1} - y^\star &\approx \left[1 - \omega(1 - \varkappa)\right] (\overline{y}^\top_t - y^\star) + \omega\, \varkappa\, y^\star\, (r_<(t) - 1), \\
    r_<(t+1) - 1 &\approx \gamma\, (r_<(t) - 1) + (\gamma - 1).
\end{align}
Since $r_<$ will decay to zero as soon as $\gamma < 1$, the transition to the fragile regime occurs at $\gamma = 1$, i.e.\
\begin{equation}\label{eq:app_sigmac}
    \sigma_c(\varkappa) = \frac{\log\varkappa}{2\sqrt{\log K}}.
\end{equation}
For $\sigma < \sigma_c$, the worst-supplier ratio recovers and the 
system returns to the steady state. 
For $\sigma > \sigma_c$, the ratio geometrically decay as $r_<(t) \sim \gamma^t$ until it becomes small 
enough that $\Pi_t < 1$, at which point the average $\overline{y}^\top_t$ 
itself begins to decay through \eqref{eq:app_avg_regime2}.

\subsection{Crash time}
\label{app:HP-crashtime}

When $\sigma > \sigma_c$, starting from the steady state, the worst 
ratio decays as $r_<(t) \sim \gamma^{-t}$ (using $\gamma > 1$ in this 
regime), while $\overline{y}^\top_t$ remains close to $y^\star$ until 
$\Pi_t$ first dips below one. From \eqref{eq:app_Pi_simplified}, 
$\Pi_t \approx \varkappa\, r_<(t) \cdot 1 \approx \varkappa\, \gamma^{-t}$ 
near the steady state, so the crossing condition $\Pi_t = 1$ gives
\begin{equation}
    \varkappa = \gamma^{t_\times} = \varkappa^{t_\times}\, e^{-2\sigma\sqrt{\log K}\, t_\times},
\end{equation}
which on taking logarithms gives
\begin{equation}\label{eq:app_crash_time}
    t_\times = \frac{\log\varkappa}{2\sigma\sqrt{\log K} - \log\varkappa} = \frac{\sigma_c}{\sigma - \sigma_c}.
\end{equation}
This is the crash time \eqref{eq:crash_time_main} of the main text. It 
diverges as $\sigma \searrow \sigma_c$, meaning that just past the 
critical line the economy can appear stable for arbitrarily long periods 
before collapsing.

\subsection{Upper bound $\sigma_{\max}$}
\label{app:HP-sigma-max}

The stability condition $\sigma < \sigma_c(\varkappa)$ derived above 
must be combined with the steady-state existence condition 
\eqref{eq:app_HS_noisy}, $\zeta_0 > \varkappa\, e^{\sigma^2}$. Taking 
logarithms, both conditions place bounds on $\log\varkappa$:
\begin{equation}
    2\sigma\sqrt{\log K} < \log\varkappa < \log\zeta_0 - {\sigma^2}.
\end{equation}
The lower bound comes from the linear-stability requirement; the upper 
bound comes from the Hawkins--Simon-type existence requirement. As 
$\sigma$ grows, the lower bound rises and the upper bound falls, and 
the interval closes at the value $\sigma_{\max}$ where they coincide
\begin{equation}
    2\sigma_{\max}\sqrt{\log K} = \log\zeta_0 - {\sigma_{\max}^2}.
\end{equation}
Solving for $\sigma_{\max}$ then yields
\begin{equation}\label{eq:app_sigma_max}
    \sigma_{\max} = 2\left(\sqrt{\log\!\big(K\, \zeta_0\big)} - \sqrt{\log K}\right),
\end{equation}
recovering Eq.~\eqref{eq:sigma_max} of the main text. For 
$\sigma > \sigma_{\max}$, no value of $\varkappa$ simultaneously 
satisfies both conditions and the economy is unstable regardless of 
inventory holdings.

\section{Rare event driven crashes}\label{app:rare-events}

\begin{figure}
    \centering
    \includegraphics[width=0.7\linewidth]{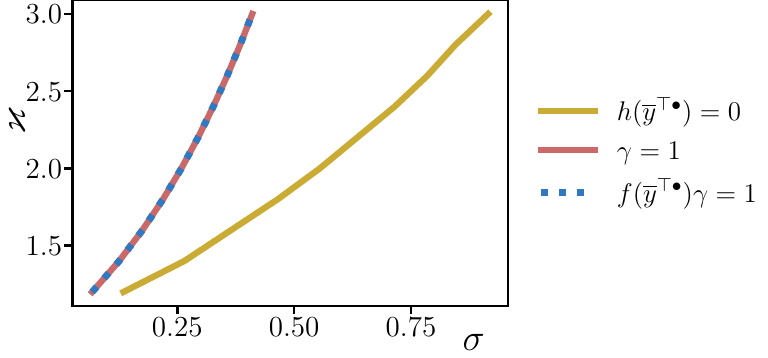}
    \caption{Critical values of $\sigma$ corresponding to the disappearance of the fixed point $h(\overline{y}^{\top \bullet})=0$ (yellow line), compared to the main text condition $\gamma=1$ (red line), which is indistinguishable from its more precise version $f(\overline{y}^{\top \bullet})\gamma=1$ (dotted blue line).}
    \label{fig:critical_curves}
\end{figure}

In the main text, we analyzed the dynamical equations in the high perishability limit using the assumption that fluctuations do not affect the equilibrium value of the production target. Could rare instantaneous events decrease this equilibrium value and shift the transition line that we computed in the main text? This is what we attempt to answer in this Appendix. 
To discuss the potential of activated events we assume that the system is initially in a stable, nearly homogeneous equilibrium state where $y_< \approx \overline{y}^\top_t \approx y^\star$ and that we will have a high fluctuation of one of the $\Xi_i$'s that will drive the productivity downwards. In this close-to-equilibrium configuration that has just been shocked, Eq. \eqref{eq:limit_prod} becomes 
\begin{align}
    y_j(t)=&\overline{y}^\top_j(t)\min\left(1, \varkappa e^{\Xi_j} \frac{\zeta y_{<}(t)}{\varkappa \overline{y}^\top_t + \zeta c}\right) \approx \overline{y}^\top_j(t)\min\left(1,\varkappa e^{\Xi_j}\right)\;,
\end{align}
since $\zeta y_{<}(t)(\varkappa \overline{y}^\top_t + \zeta c)^{-1}=f(y^\star)=1$.
In the case of a shock at time $t$, the realised production is dominated by the second term which is such that $\varkappa e^{\Xi_j} < 1$. The probability $p(\varkappa)$ for such a shock to happens reads 
\begin{align}
p(\varkappa)=&\int_{-\infty}^{-\log(\varkappa)} \frac{ e^{- \frac{\xi^2}{4 \sigma^2}}}{\sqrt{4 \pi \sigma^2}} =\frac{1}{2}{\rm{Erfc}}\left[\frac{\log(\varkappa)}{2 \sigma}\right]
\end{align}
The number of firms infected by such adverse shocks is thus $Np(\varkappa)$. Each such firm $j$ leads to a drop of inputs for the $K$ firms that are clients of $j$. For each of these client firms $i\in \mathcal{C}(j)$, we have 
\begin{align}
\overline{x}_{ij} \approx \varkappa y_j \approx \varkappa^2 y^\star e^{\Xi_j}
\end{align}
and thus, by definition of firm $j$, $\overline{x}_{ij} < \varkappa y^\star$ thereby entailing a bottleneck as well for its customer $i$.
If furthermore the stronger condition  $\varkappa^2 e^{\Xi_j} < 1$ holds, firm $i$ will update its production target downwards in the next time round
\begin{align}
y^\top_i(t=1)= (1 - \omega) y^\star + \omega \varkappa^2 y^\star e^{\Xi_j}\;.
\end{align}
This is the case for $Np(\varkappa^2) \times K$ firms, so the average production target is modified according to
\begin{align}
\overline{y}^\top_{t+1}= y^\star  \left(1  + \omega p(\varkappa^2) K [ \varkappa^2  \langle e^{\Xi_j} \rangle_< - 1]\right),
\end{align}
where the $\langle \cdot\rangle_<$ means the average of $e^{\Xi_j}$ conditioned to $\varkappa^2 e^{\Xi_j} < 1$. This is given by 
\begin{align}
p(\varkappa^2)\langle e^{\Xi_j} \rangle_< = \int_{2\log \varkappa}^\infty {\rm d}\xi \, \frac{ e^{-\xi - \frac{\xi^2}{4 \sigma^2}}}{\sqrt{4 \pi \sigma^2}} = \frac{1}{2}e^{\sigma^2}{\rm{Erfc}}\left[\sigma + \frac{\log(\varkappa)}{\sigma}\right]\;.
\end{align}
In addition, we have that 
\begin{align}
\label{eq:pb2}
    p(\varkappa^2)=\frac{1}{2}{\rm{Erfc}}\left[\frac{\log(\varkappa)}{\sigma}\right]\;,
\end{align}
so that, at the next time step, the planned production is affected according to 
\begin{align}
\overline{y}^\top_{t+1}= y^\star  \left(1  + \omega g\left(K,\varkappa,\sigma\right)\right),
\end{align}
where $g(K,\varkappa,\sigma)$ can be analytically and explicitly computed from the above expressions. As soon as $g(K,b,\sigma)<0$, the economy is prone to crashes. 
We now generalize this argument at an arbitrary time $t$ for which the average production target is $\overline{ y}_t$. 
In the case of a strong adverse shock to firm $j$ that was in a ``normal'' situation previously, i.e. $y^\top_j \approx \overline{ y}_t$, the impact on its $K$ customers $i$ is
\begin{align}
\label{eq:avalanche_plan_prod_before_av}
y^\top_i(t+1)= (1 - \omega) y^\top_i(t) + \omega (c+\frac{\varkappa}{\zeta}\overline{y}^\top_t) \min\left(1,\varkappa^2 e^{\Xi_j} \frac{y^\top_i(t)\overline{y}^\top_t}{(c+\frac{\varkappa}{\zeta}\overline{y}^\top_t)^2}\right).
\end{align}
The distinction between the two cases within the $\min$ function is now determined by the condition 
\[
\varkappa^2 f_t^2 e^{\Xi_j}  < 1, \qquad f_t:= \frac{\zeta \overline{y}^\top_t}{\zeta c+\varkappa\overline{y}^\top_t},
\]
which is increasingly easier to satisfy when $\overline{y}^\top_t$ decreases -- we recall that $f_t$ is a decreasing function of $\overline{y}^\top_t$. Cascading production failures (or ``avalanche'') may thus speed up with time. 
Defining $p(\varkappa^2f_t^2)$ as the probability that $\varkappa^2 f_t^2 e^{\Xi_j}  < 1$, we can now average \eqref{eq:avalanche_plan_prod_before_av} over all firms to obtain the evolution of $ \overline{y}^\top$.
We note that $Np(\varkappa^2f_t^2)K$ firms will have the factor $\varkappa^2f_t^2 e^{\Xi_j}$ chosen from the $\min$ in \eqref{eq:avalanche_plan_prod_before_av}, while $N-Np(\varkappa^2f_t^2)K$ other firms will have the $1$ chosen inside the $\min$ operation. Hence, we get 
\begin{align}
\overline{y}^\top_{t+1}= (1 - \omega) \overline{y}^\top_t + \omega (c+\frac{\varkappa}{\zeta}\overline{y}^\top_t)(1-K p(\varkappa^2f_t^2)) + \omega (c+\frac{\varkappa}{\zeta}\overline{y}^\top_t)K\varkappa^2 f_t^2 p(\varkappa^2f_t^2) \langle e^{\Xi_j} \rangle_<,
\end{align}
where the $\langle \cdot\rangle_<$ means the average of $e^{\Xi_j}$ conditioned to $\varkappa^2 f_t^2 e^{\Xi_j} < 1$.
We then compute that
\begin{align}
    p(\varkappa^2f_t^2)\langle e^{\Xi_j} \rangle_< = \int_{2\log(\varkappa f_t)}^\infty {\rm d}\xi \, \frac{ e^{-\xi - \frac{\xi^2}{4 \sigma^2}}}{\sqrt{4 \pi \sigma^2}}=\frac{1}{2}e^{\sigma^2}{\rm Erfc}\left[\frac{\log(\varkappa f_t)}{\sigma}+\sigma\right]\;,
\end{align}
while
\begin{align}
p(\varkappa^2 f_t^2) =\frac{1}{2}{\rm Erfc}\left[\frac{\log(\varkappa f_t)}{\sigma} \right] \;.
\end{align}
Hence, the update rule for $\overline{y}^\top_{t}$ becomes 
\begin{align}
\label{eq:avalanche_closed_production}
\overline{y}^\top_{t+1} - \overline{y}^\top_t = \omega 
h( \overline{y}^\top_t)
\end{align}
with
\begin{align}
h( \overline{y}^\top_t):=
(c+\frac{\varkappa}{\zeta}\overline{y}^\top_t)\left(1-\frac{K}{2}{\rm Erfc}\left[\frac{\log(\varkappa f_t)}{\sigma} \right] + \frac{K}{2}\varkappa^2f_t^2e^{\sigma^2}{\rm Erfc}\left[\frac{\log(\varkappa f_t)}{\sigma}+\sigma\right] \right) -  \overline{y}^\top_t\;.
\end{align}
Eq \eqref{eq:avalanche_closed_production} is a closed evolution equation for the planned production $\overline{y}^\top$, with a fixed point $\overline{y}^{\top \bullet}$ such that $h(\overline{y}^{\top \bullet})=0$. When $\sigma$ is small enough, $\overline{y}^{\top \bullet}$ is close to the equilibrium value $y^\star$ in the main text. This fixed point disappears in a region outside the stability determined in the main text (see Fig. \ref{fig:critical_curves}), so this instability channel can be neglected. Throughout the stability region, we find that $f(\overline{y}^{\top \bullet})\gamma \approx \gamma$; therefore the simple criterion $\gamma=1$ derived in the main text is a very good approximation for the instability, see Fig. \ref{fig:critical_curves}. 

\section{Average production shortfall in a Brownian framework}
\label{app:shortfall}

This appendix derives, in a self-contained way, the closed-form expression for the expected exceedance integral quoted in the main text. Because the techniques (Brownian motion with drift, Feynman–Kac representations, Green's functions of killed diffusions, Itô excursion theory) are not standard in economics, we provide detailed intuition alongside the technical steps.

\subsection*{Average cost of an excursion}
We consider a Brownian motion with negative drift,
\begin{align}
\label{eq:brownian_drift}
dB_t = -\eta + dW_t
\end{align}
where $W_t$ is a standard Wiener process and $\eta$ is a positive drift. Denoting by $\tau_1$ the first passage time to the origin of a realization $\{B_t\}$, we want to derive the average cost of an excursion starting at $\epsilon$ above the origin, namely
\begin{align}
    h(\epsilon)= \mathds{E}_{B_t}\!\left[\int_0^{\tau_1}f(B_t)\,dt \bigg| B_0=\epsilon \right],
\end{align}
where the average $\mathds{E}$ runs over all realizations of \eqref{eq:brownian_drift} and $f$ is a cost function. 
Note that since $\eta >0$, each realization $\{B_t\}$ has a well defined $\tau_1$.
We define the probability of transitioning from $B_0=\epsilon$ to $B_t=y$ as $p(y,t|\epsilon,0)$.
Since we only consider paths that have not passed the origin yet, the boundary condition for this transition probability is $p(0,t|\epsilon,0)=0$ at any $t$.
The average cost at time $t$ is then obtained as 
\begin{align}
    \mathcal{C}(\epsilon,t)=\int_0^{\infty} f(y) p(y,t|\epsilon,0) dy\;.     
\end{align}
In addition, the transition probability $p(y,t|\epsilon,0)$ evolves according to the backward Fokker-Planck as 
\begin{align}
\label{eq:backward_FP}
    \partial_t p(y,t|\epsilon,0) = -\eta \partial_{\epsilon}p(y,t|\epsilon,0) + \frac{1}{2} \partial_{\epsilon \epsilon}p(y,t|\epsilon,0)\;.
\end{align}
Multiplying \eqref{eq:backward_FP} by $f(y)$ and integrating over $y$ then yields the evolution equation for the cost $\mathcal{C}(\epsilon,t)$ as 
\begin{align}
\label{eq:cost_evolution}
    \partial_t \mathcal{C}(\epsilon,t) = -\eta \partial_{\epsilon} \mathcal{C}(\epsilon,t) + \frac{1}{2} \partial_{\epsilon\epsilon} \mathcal{C}(\epsilon,t)\;.
\end{align} 
The quantity we aim at deriving, $h(\epsilon)$, is the total cost over trajectories, namely $h(\epsilon)=\int_0^{\infty}\mathcal{C}(\epsilon,t)dt$.
Integrating \eqref{eq:cost_evolution} between $t=0$ and $t=\infty$ yields 
\begin{align}
    \mathcal{C}(\epsilon,\infty)-\mathcal{C}(\epsilon,0)=-\eta \partial_{\epsilon} h(\epsilon) + \frac{1}{2} \partial_{\epsilon\epsilon} h(\epsilon)\;.
\end{align}
Since $\eta$ is negative, a realization $\{B_t\}$ has to cross the origin at some finite time and $p(y,\infty|\epsilon,0)=0$, entailing that $\mathcal{C}(\epsilon,\infty)=0$.
Furthermore, $p(y,0|\epsilon,0)=\delta(y-\epsilon)$ so that $\mathcal{C}(\epsilon,0)=f(\epsilon)$ and we finally obtain the equation verified by $h(\epsilon)$ as 
\begin{align}
\label{eq:final_cost_eq}
    -f(\epsilon )=-\eta \partial_{\epsilon} h(\epsilon) + \frac{1}{2} \partial_{\epsilon\epsilon} h(\epsilon)\;.
\end{align}
Its corresponding boundary condition is $h(0)=0$ since when the Brownian motion starts at $\epsilon=0$ it immediately escapes.
Since we are interested in paths starting at the origin before returning to the origin for the first time, the object we aim to derive is the excursion-measure limit
\begin{align}
\Phi = \lim_{\epsilon \to 0^+}\frac{h(\epsilon)}{\epsilon},
\end{align}
which captures the expected cost per unit of initial displacement from the origin, in the limit where the excursion starts from $0^+$.
We now set up to derive $h(\epsilon)$ for our specific cost function.

\subsection*{The Green function of the excursion's cost}
Rather than solving \eqref{eq:final_cost_eq} for a specific cost function $f$, we generically solve it by introducing the Green function $G(\epsilon,y)$ satisfying
\[
\left[ - \eta \partial_\epsilon + \frac{1}{2}\partial_{\epsilon\epsilon}\right]G(\epsilon,y) = -\delta(\epsilon-y)
\]
Once $G$ is known, the solution to \eqref{eq:final_cost_eq} is given by the convolution
\begin{align}
\label{eq:sol_excursion}
h(\epsilon) = \int_0^{\infty} f(y)\,G(\epsilon,y)\,dy.
\end{align}
Since $h(0)=0$, $G$ must further satisfy $G(0,y) = 0$.
Probabilistically, $G(\epsilon,y)\,dy$ is the expected time, starting from $\epsilon$, that $B_t$ spends in the infinitesimal interval $[y,y+dy]$ before hitting the origin.
The general solution to the homogeneous equation
\[
\tfrac12 u''(\epsilon) - \eta\,u'(\epsilon)=0,
\]
is $u(\epsilon) = A + B\,e^{2\eta\epsilon}$. 
For $\epsilon\ne y$, $G$ thus reads
\begin{align}
G(\epsilon,y)=
    \begin{cases}
A_1(y) + B_1(y)\,e^{2\eta\epsilon}, & \text{when }0<\epsilon<y,\\
A_2(y) + B_2(y)\,e^{2\eta\epsilon}, & \text{when }\epsilon>y>0.
\end{cases}\;
\end{align}
The coefficients $A_1$, $B_1$, $A_2$ and $B_2$ are determined by
\begin{enumerate}
\item $G(0,y) = 0$ (absorbing boundary at the origin);
\item $G(\epsilon,y)$ bounded as $\epsilon\to\infty$ (since $\eta>0$, the drift pushes the process toward zero);
\item continuity of $G$ at $\epsilon = y$;
\item a jump of $-2$ in $\partial_\epsilon G$ at $\epsilon = y$ (to produce the Dirac source).
\end{enumerate}
Straightforward algebra gives the final expression\footnote{A quick way to derive this: on $\epsilon < y$, impose (1) to get $G\propto e^{2\eta\epsilon}-1$; on $\epsilon>y$, impose (2) to get $G\propto \text{const}$; then match using (3)--(4).}
\begin{align}
G(\epsilon,y)=
\begin{cases}
\dfrac{e^{-2\eta y}\bigl(e^{2\eta\epsilon}-1\bigr)}{\eta}, & 0<\epsilon<y,\\[10pt]
\dfrac{1-e^{-2\eta y}}{\eta}, & \epsilon>y>0.
\end{cases}\;
\end{align}
Integrating $G$ over all $y$ gives the average first-passage time as
\[
\int_0^\infty G(\epsilon,y)\,dy \;=\; \frac{\epsilon}{\eta} \;=\; \mathds{E}_{B_t}[\tau_1],
\]
which corresponds to the classical result for a Brownian motion with negative drift starting at $\epsilon$.

\subsection*{Deriving the excursion's cost}
We now derive the average cost of excursion for our specific case where $f$ reads
\begin{align}
f(y) = 1 - e^{-(y-X)_+} = \begin{cases}
0, & 0\le y\le X,\\
1 - e^{X-y}, & y>X.
\end{cases}
\end{align}
Because $f(y)=0$ on $y\le X$, $h(\epsilon)$ is given by the integral \eqref{eq:sol_excursion} as
\[
h(\epsilon) \;=\; \int_X^{\infty} \bigl(1 - e^{X-y}\bigr)\,G(\epsilon,y)\,dy.
\]
Since the initial condition $\epsilon$ is small (ultimately we send $\epsilon\to 0$), we are in the regime $0<\epsilon<X$, and thus we use the first branch of $G$
\[
G(\epsilon,y) = \frac{e^{-2\eta y}\,(e^{2\eta\epsilon}-1)}{\eta}, \qquad \text{for } y>\epsilon\;,
\]
giving
\begin{align}
h(\epsilon) \;=\; \frac{e^{2\eta\epsilon}-1}{\eta}\int_X^{\infty}\bigl(1-e^{X-y}\bigr)\,e^{-2\eta y}\,dy = \frac{\bigl(e^{2\eta\epsilon}-1\bigr)\,e^{-2\eta X}}{2\eta^2\,(2\eta+1)}
\end{align}
The excursion measure $\Phi$ is then obtained as 
\begin{align}
\label{eq:excursion_measure_final}
    \Phi =\lim_{\epsilon\to 0^+}\frac{f(\epsilon)}{\epsilon} = \frac{e^{-2\eta X}}{\eta\,(2\eta+1)}
\end{align}

\subsection*{Rescaling to the parameters of the main text}
In the main text, the Brownian motion carries a diffusion coefficient $\Gamma>0$:
\[
dX_t = \Gamma\,dW_t - \eta\,dt.
\]
This amounts to a rescaling $\widehat B_t := X_t/\Gamma$, which evolves as 
\[
d\widehat B_t = dW_t - (\eta/\Gamma)\,dt,
\]
so $\widehat B$ is a standard drifted Brownian motion with drift $-\widehat\eta := -\eta/\Gamma$, starting from $\widehat\epsilon := \epsilon/\Gamma$. and the threshold transforms to $\widehat X := X/\Gamma$.
Taking into account that the threshold transforms to $\widehat X := X/\Gamma$, formula \eqref{eq:excursion_measure_final} derived above yields, in hatted variables,
\[
\widehat\Phi(\widehat X) \;=\; \frac{e^{-2\widehat\eta\,\widehat X}}{\widehat\eta\,(2\widehat\eta+1)}
\;=\;\frac{e^{-2\eta X/\Gamma^2}}{(\eta/\Gamma)\,(2\eta/\Gamma+1)}.
\]
Accounting for the rescaling of $\epsilon$ (since $\Phi$ is a per-unit-$\epsilon$ prefactor, one factor of $\Gamma$ enters), the unscaled excursion prefactor in the original variables is
\begin{align}
\Phi \;=\; \frac{\Gamma^2\,e^{-2\eta X/\Gamma^2}}{\eta\,(2\eta+\Gamma^2)}\;,
\end{align}
which matches the formula quoted in the main text.

\end{document}